\documentstyle[prb,aps]{revtex}
\begin{document}
\input psfig
\draft

\title{Resonant tunneling through ultrasmall quantum dots: zero-bias 
anomalies, magnetic field dependence, and boson-assisted transport}

\author{J\"urgen K\"onig$^1$, J\"org Schmid$^1$, Herbert Schoeller$^1$, and 
Gerd Sch\"on$^{1,2}$}

\address{
$^1$ Institut f\"ur Theoretische Festk\"orperphysik, Universit\"at Karlsruhe, 
76128 Karlsruhe, Germany \\
$^2$ Department of Technical Physics, Helsinki University of Technology, 
02150 Espoo, Finland}

\date{\today}

\maketitle

\begin{abstract}
We study resonant tunneling through a single-level quantum dot in the 
presence of strong Coulomb repulsion beyond the perturbative regime.
The level is either spin-degenerate or can be split by a magnetic field.
We, furthermore, discuss the influence of a bosonic environment.
Using a real-time diagrammatic formulation we calculate transition rates, the 
spectral density and the nonlinear $I-V$ characteristic.
The spectral density shows a multiplet of Kondo peaks split by the transport 
voltage and the boson frequencies, and shifted by the magnetic field.
This leads to zero-bias anomalies in the differential conductance, which agree
well with recent experimental results for the electron transport through 
single-charge traps.
Furthermore, we predict that the sign of the zero-bias anomaly depends on the 
level position relative to the Fermi level of the leads. 
\end{abstract}

\section{Introduction}
The experimental study of tunneling through zero-dimensional states in quantum
dots with high charging energies has received recently considerable interest 
\cite{Kou-rev,Gol-Cun,Gue-etal,Joh-etal,Fox-etal,Weiss,Ral-Bla-Tin}. 
Theoretical studies cover the classical regime (high temperatures) 
\cite{Ave-Kor-Lik,Bee,Wei,Bru-HS,Gla-Mat} as well as the quantum-mechanical 
regime (low temperatures) 
\cite{Her-Dav-Wil,Mei-Win-Lee,Koe-HS-Sch2,Gro1,Ng-Lee,Gla-Rai,Kaw,Ng1,YMF,Het-HS,Bru-Faz-HS}. 
In the latter case, Coulomb blockade and resonant-tunneling phenomena together
with nonequilibrium generalizations of the Kondo effect are expected to occur.
This leads to zero-bias anomalies in the differential conductance, which 
recently have been observed by Ralph \& Buhrman \cite{Ral-Buh1}. 
In this article, we present a new real-time diagrammatic approach to describe 
resonant tunneling at low temperatures and compare our results to the latter 
experiment.

The experiments for Coulomb blockade phenomena in zero-dimensional systems are
usually performed in double-barrier resonant-tunneling structures
\cite{Gol-Cun,Gue-etal}, split-gate quantum-dot devices 
\cite{Joh-etal,Fox-etal,Weiss}, quantum point-contacts with single-charge trap
states \cite{Ral-Buh1} and quite recently also in ultrasmall metallic tunnel 
junctions \cite{Ral-Bla-Tin} with Al particles of diameter below $10nm$.
In the latter experiment, the level spacing is of order $0.5meV$ which is 
comparable to the Coulomb charging energy in usual quantum dots. 
Therefore, the quantum dot is described by the nonequilibrium Anderson model 
where the energy level $\epsilon_\sigma$ (with spin label $\sigma$) is coupled
via tunneling barriers to two electron reservoirs with different 
electrochemical potentials $\mu_L$ and $\mu_R$. 
The charging energy is described by a strong on-site Coulomb repulsion $U$ 
which suppresses double occupancy of the dot level. 
In equilibrium, it is well known from the theory of strongly correlated 
fermions \cite{Bic} that the spectral density of the dot can exhibit a Kondo 
resonance at the Fermi level. 
It occurs for a low lying level $\epsilon_\sigma -\mu_{L,R} < -\Gamma$ and 
weak Zeeman splitting $|\epsilon_\sigma-\epsilon_{\bar{\sigma}}| < \Gamma$,  
where  $\Gamma/2$ is the level width in the noninteracting case, and provided 
that temperature is lower than the Kondo temperature \cite{Bic,Hal}
$T_K=1/2(U\Gamma)^{1/2}\exp{[\pi\epsilon (\epsilon+U)/(\Gamma U)]}$.
Since the weight of the equilibrium spectral density at the Fermi level is 
proportional to the linear conductance, an enhancement of the latter due to 
Kondo-assisted tunneling was predicted \cite{Ng-Lee,Gla-Rai}. 
Typical values for quantum dots are $U\sim 1meV$ and $\Gamma\sim 50\mu eV$ 
which, for $\epsilon\sim -\Gamma$, yield a Kondo temperature of the order 
$T_K\sim 50mK$. 
Due to heating effects such temperatures are still hard to realize in 
realistic dots.
A more pronounced feature was found for the nonlinear conductance which shows 
a zero-bias maximum even for temperatures above the Kondo temperature 
\cite{Her-Dav-Wil,Mei-Win-Lee}. 
At zero magnetic field the spectral density of each spin channel exhibits
a Kondo resonance at each of the chemical potentials.
An applied magnetic field causes the Kondo peaks to shift from the chemical 
potential by the Zeeman energy, in opposite directions for the different spin
channels.
As a consequence, Kondo-assisted tunneling can only occur if the bias voltage 
exceeds the Zeeman energy.
Therefore, the zero-bias anomaly is split by an applied magnetic field 
\cite{Mei-Win-Lee}. 
These features have been observed experimentally by Ralph \& Buhrman 
\cite{Ral-Buh1}. 
They measured the differential conductance through single-charge traps in a 
metallic quantum point-contact. 
Although this system does not allow a controlled variation of the level 
position, the appearance of a zero-bias maximum with a peak height varying 
logarithmically with temperature clearly demonstrates the mechanism of 
Kondo-assisted tunneling. 
However, the detailed comparison of the line shape in the experiment and the 
existing theory showed significant deviations.

In this paper, we will describe a new approach to calculate nonequilibrium
properties of strongly correlated mesoscopic systems coupled to fermionic 
or bosonic baths via particle and energy exchange. 
It consists in a real-time diagrammatic approach closely related to 
path-integral methods formulated in connection with dissipation 
\cite{Fey-Ver,Leg-etal,Wei-Buch} or tunneling in metallic junctions 
\cite{Eck-Sch-Amb,Sch-Zai,Sch-Uebersicht}. 
One of the difficulties of the present problem lies in the fact that we have 
to account for the Coulomb interaction in a nonperturbative way.
Therefore, the standard many-body diagrammatic approaches are not sufficient
(since Wick's theorem cannot be applied naively).
We circumvent the problem by keeping track explicitly of the time evolution of
the density matrix of the dot and tracing out only the bath degrees of 
freedom, which are assumed to be in equilibrium. 
The final diagrammatic language is set up by an expansion in the coupling to 
the fermionic reservoirs whereas the strong correlations on the local system 
are exactly taken into account.
The basic step is the calculation of transition and current rates between 
different states of the dot. 
We present an exact expression for these rates as the sum over all irreducible
diagrams. 
The transition rate is used to set up a formally exact Master equation from 
which the time-dependent probability distribution for the dot can be 
calculated.
The current rate is generally not identical to the transition rate since it 
contains also the number of particles transferred to the reservoir where the 
current is calculated. 
This number can take arbitrary values if one considers all higher order 
processes.
The occupation probabilities multiplied with the current rates are used to 
calculate the current flowing through the system.
In earlier publications, we have presented this technique in connection with 
tunneling through a metallic island with a continuum of states \cite{HS-Sch} 
and demonstrated the equivalence to path-integral methods \cite{Koe-HS-Sch1}.
There we used an approximation for the rate to sum up ``inelastic resonant
tunneling'' processes to arbitrary order where different electrons tunnel 
coherently back and forth between the island and the reservoirs. 
Here, we apply an equivalent approximation to describe resonant tunneling 
between metallic leads through an ultrasmall quantum dot with a single level. 
An important advantage of our approach is that we can solve the 
noninteracting limit exactly and can control systematically if this limit is 
contained within a given approximation for the correlated case. 
The theory is current conserving and can be used for the calculation of 
correlation functions or Green's functions as well.

For the case where the dot level is $M$-fold degenerate we recover for 
$M\ge 2$ and a low lying dot level a Kondo peak in the spectral function.
An applied transport voltage leads to a splitting of the Kondo peak (at 
$\mu_\alpha$ where $\alpha$ denotes the lead), which results in a zero-bias 
anomaly in the differential conductance, such that the conductance has a 
maximum at $V=0$.
On the other hand, if the dot level lies above the Fermi levels of the leads 
$\mu_\alpha$ we predict \cite{Koe-HS-Sch2} a zero-bias anomaly in the 
conductance which  has a minimum  at $V=0$.

Several extensions will be considered.
We study the case where the (spin-) degeneracy of the dot level is lifted,
e.g., by an applied magnetic field.
In this case the Kondo peaks of both spin channels move apart from each other 
by the level spacing $\epsilon_{\sigma}-\epsilon_{\bar{\sigma}}$, and 
Kondo-assisted tunneling sets in only at transport voltages $eV$ exceeding 
this splitting.
The calculated conductance agrees well with the experimental results of 
Ref.~\onlinecite{Ral-Buh1}.

We, furthermore, account for inelastic interactions with bosonic modes coupled
to the dot. 
They describe applied time-dependent fields, interaction with phonons, or the 
fluctuations in the electrodynamic environment. 
The investigation of this field has started only recently in connection with 
transport through interacting quantum dots.
The classical regime ($\Gamma \gg T$) has been analyzed for time-dependent
fields \cite{Bru-HS} and bosonic environments \cite{Sta}. 
Photon- and boson-assisted tunneling leads here to resonant side peaks in the 
Coulomb oscillations which can be used to analyze the complete excitation 
spectrum of the dot. 
The results agree well with experiments \cite{Delft-Diplom}. 
In the Kondo regime it has been found \cite{Het-HS} that time-dependent 
perturbations split the Kondo resonances which leads to satellite anomalies in
the differential conductance and offers the possibility to realize pump 
effects which are purely based on the presence of Kondo resonances.
The linear AC-conductance has been analyzed in Ref.~\onlinecite{Ng2}.
In this paper, we will investigate the influence of an external bosonic field 
on transport phenomena through ultrasmall quantum dots at low temperatures and
small boson frequencies (compared to $\Gamma$). 
The emission and absorption of bosons causes additional Kondo singularities, 
for a one-mode field at $\mu_\alpha+n\omega_B$, where $n=\pm 1,\pm 2,\ldots$. 
Again, these resonances lead to corresponding anomalies in the differential 
conductance, which are inverted if the level position of the dot is moved 
through the Fermi energy.

\section{Hamiltonian}

\begin{figure}
\centerline{\psfig{figure=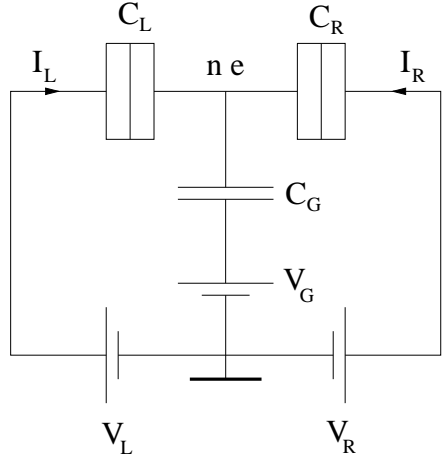,height=6cm}}
\caption{The SET transistor.}
\label{Fig.transistor}
\end{figure}

We concentrate here on a dot containing only one energy level 
$\epsilon_\sigma^{(0)}$ with degeneracy $M$.
In a magnetic field, due to the Zeeman energy the level position is spin 
dependent. 
The dot is connected via high tunneling barriers to two large noninteracting 
reservoirs and coupled capacitively to an external gate voltage. 
The model Hamiltonian of this ``single-electron transistor''
(see Fig.~\ref{Fig.transistor}) is
\begin{equation}
	H=H_0+H_T=H_D+\sum_{\alpha=L,R}H_{\alpha}+H_T \, .
\label{HH}
\end{equation}
Here, $H_D$ is the Hamiltonian for the dot. 
It includes the Coulomb interaction of the dot electrons, which is described 
within the capacitance model of a single-electron transistor by the 
capacitances $C_L$ and $C_R$ for the left and right tunnel junction and $C_G$ 
for the gate.
Furthermore, the dot electrons are coupled to bosonic modes $\omega_q$ with 
electron-boson coupling $g_q$.
Thus, $H_D$ reads
\begin{equation}
	H_D=\sum_{\sigma}\epsilon_\sigma^{(0)} n_{\sigma}+
	E_C\left( \hat{N} - n_G \right) ^2+\sum_q \omega_q d_q^\dagger d_q
	+\hat{N}\sum_q g_q (d_q + d_q^\dagger)\;. 
\label{HDot}
\end{equation}
(Throughout this work, we set $\hbar=k=1$ and use the convention $e>0$.)
The particle number on the dot with spin $\sigma$ is 
$n_\sigma=c^\dagger_\sigma c_\sigma$, and $\hat{N} =\sum_{\sigma} n_{\sigma}$.
The scale of the charging energy is provided by $E_C\equiv e^2/ 2C$, where
$C=C_L+C_R+C_G$ is the total capacitance of the system.
The transistor can be tuned by the gate voltage $V_G$ via 
$e n_G= C_LV_L+C_RV_R+C_GV_G$. 
We remark here, that $H$ is invariant under a global shift of all energies.
Therefore, we can always choose symmetric bias, $V_L=-V_R=V/2$.

The next term in Eq.~(\ref{HH}), 
$H_{\alpha}=\sum_{k\sigma}\epsilon_{k\alpha}a^\dagger_{k\sigma\alpha}
a_{k\sigma\alpha}$, describes the reservoir $\alpha$ of noninteracting 
electrons in the leads. 
Finally, the dot is coupled via tunnel barriers to the left and right lead.
This coupling is described by the tunnel Hamiltonian
\begin{equation}
	H_T=\sum_{k\sigma\alpha}(T^\alpha_{k} a^\dagger_{k\sigma\alpha}
	c_\sigma + h.c.).
\end{equation}

The bosonic modes can represent interaction with phonons 
\cite{Win-Jac-Wil,Gla-She,Jon} or fluctuations of the electrodynamic 
environment \cite{Devoret-etal,Odi-etal,Flens-etal,Ima-Pon-Ave}
very similar to the Caldeira-Leggett model\cite{Cal-Leg}.
For our theory no assumption is needed for the specific kind of the modes 
$\omega_q$ and the couplings $g_q$. 
In this way we are able to present a general result for the current which 
shows the influence of inelastic interactions for an arbitrary environment.

The capacitive model is equivalent to the Anderson Hamiltonian, which we 
obtain by defining the interaction $U_0=2E_C$ and shifting the level position 
$\epsilon_\sigma^{(0)}+2E_C- eV_G C_G/C -a_ceV/2 \rightarrow 
\epsilon^{(0)}_\sigma(V,V_G)$.
The asymmetry factor $a_c=(C_L-C_R)/C$ accounts for a different capacitive 
coupling of the left and right lead.
We see here, that the effective level position in the Anderson model depends 
on the gate voltage, as well as, due to $a_c$, on the transport voltage.
The dot is then described by
\begin{equation}
	H_D=\sum_{\sigma}\epsilon_\sigma^{(0)}(V,V_G) \, n_{\sigma}+
        U_0\sum_{\sigma < \sigma'} n_\sigma n_{\sigma'}
	+\sum_q \omega_q d_q^\dagger d_q
	+\hat{N}\sum_q g_q (d_q + d_q^\dagger) \; .
\end{equation}

An unitary transformation \cite{Mah} with $V=\exp(-i\hat{N}\varphi)$ and 
$\varphi=i\sum_q (g_q/\omega_q)(d_q^\dagger-d_q)$ yields 
$\bar{H}=VHV^{-1}=\bar{H}_0+\bar{H}_T$, where $\bar{H}_0=H_R+\bar{H}_D$,
\begin{equation}\label{dot}
	\bar{H}_D=\sum_{\sigma}\epsilon_{\sigma}n_{\sigma}
	+U\sum_{\sigma<\sigma'}n_\sigma n_{\sigma'}+\sum_q \omega_q 
	d^\dagger_q d_q
\end{equation} 
and 
\begin{equation}\label{tunnel}
	\bar{H}_T=\sum_{k\sigma\alpha}(T^\alpha_{k} 
	a^\dagger_{k\sigma\alpha}c_\sigma e^{i\varphi}+h.c.)\, .
\end{equation}
The electron-boson interaction renormalizes the level position and the Coulomb
repulsion, $\epsilon_{\sigma}=\epsilon_\sigma^{(0)}-\sum_q {g_q^2/\omega_q}$ 
and $U=U_0-2\sum_q{g_q^2/\omega_q}$, and the tunneling term acquires phase 
factors $e^{\pm i\varphi}$.

For strong Coulomb repulsion $U$ we restrict ourselves to states with $N=0,1$.
In lowest order perturbation theory the rates for tunneling in and out of the 
dot to reservoir $\alpha$ are 
\begin{equation}\label{gamma}
	2\pi \gamma_\alpha^\pm(E)=2 \pi \int dE' \bar{\gamma}^\pm_\alpha (E')
	P^\pm (E-E') \, ,
\end{equation}
where $2\pi \bar{\gamma}_\alpha^\pm(E)= \Gamma_\alpha(E)f^\pm_\alpha(E)$ 
is the classical rate without bosons, 
$\Gamma_\alpha (E)=2\pi\sum_k |T^\alpha_k|^2 \delta (E-\epsilon_{k\alpha})$, 
and $f_\alpha^+(E)$ is the Fermi distribution of reservoir $\alpha$ with
electrochemical potential $\mu_\alpha$, while 
$f_\alpha^-(E)=1-f^+_\alpha (E)$.
Finally,
\begin{equation}\label{Ppm}
	P^\pm (E)={1\over 2\pi}\int dt \, e^{iEt} <e^{i\varphi (0)}
	e^{-i\varphi (\pm t)}>_0
\end{equation}
describes the probability that an electron absorbs ($P^+$) or emits ($P^-$) 
the boson energy $E$.\cite{Devoret-etal,Odi-etal,Flens-etal}
The expectation value is taken with the free boson Hamiltonian. 
These probabilities satisfy the condition of detailed balance \cite{Ing-Naz}
\begin{equation}\label{detailed balance}
	P^-(E) = P^+(-E) = e^{\beta E} P^+(E)
\end{equation}
The classical rates combined with a master equation are sufficient in the 
perturbative regime $\Gamma = \sum_{\alpha} \Gamma_{\alpha} \ll T$.\cite{Sta}

In order to go beyond the perturbative regime, we need a nonperturbative 
treatment of the tunneling, where quantum fluctuations yield finite life-time 
broadening and renormalization effects of the dot levels. 
As an illustration we first assume that (for $B=0$) the broadening is given by
the sum of the classical transition rates, Eq.~(\ref{gamma}).
Using Kramers-Kronig we deduce the renormalization and obtain for the 
self-energy
\begin{equation}\label{sigma}
	\sigma(E)=\int dE' \, {M \gamma^+ (E')
	+\gamma^-(E')\over E-E'+i0^+}.
\end{equation}
where $\gamma^\pm=\sum_\alpha\gamma_\alpha^\pm$.
The aim of the present letter is to test and extend this simple physical 
picture within a systematic and conserving theory for all Green's functions 
and the current.
To achieve this we  use a real-time technique developed in 
Ref.~\onlinecite{HS-Sch,Koe-HS-Sch1,Koe-HS-Sch2} which provides a natural 
generalization of the classical and cotunneling theory to the physics of 
resonant tunneling.

\section{Diagrammatic technique}

A quantum-statistical expectation value of an operator $A$ at time $t$ is 
given by
\begin{equation}
	\langle  A(t)\rangle = \mbox{tr}[\rho_0  A(t)_{\bar{H}}],
\end{equation}
where $A(t)_{\bar{H}}=\exp [i\bar{H}(t-t_0)] A\exp [-i\bar{H}(t-t_0)]$ is the 
operator in  Heisenberg picture with respect to the initial time $t_0$.
Permutation under the trace yields $\langle A(t)\rangle = \mbox{tr}[\rho(t)A]$
with $A$ in Schr\"odinger picture. 
The density matrix $\rho (t)$ evolves in time via 
$\rho (t) = e^{-i\bar{H}(t-t_0)}\rho(t_0)e^{i\bar{H}(t-t_0)}$.
We assume that the initial density matrix $\rho_0=\rho (t_0)$ factorizes into
parts for the dot electrons, the bosons, and the leads:
\begin{equation}
	\rho_0=\rho_0^D\rho_0^B\prod_{\alpha}\rho_0^{\alpha}.
\end{equation}
The leads are treated as large equilibrium reservoirs with fixed 
electrochemical potentials $\mu_{\alpha}=-eV_{\alpha}$.
Therefore, we describe the electrons in the leads by Fermi functions 
$f_{\alpha}(E)$ and the density matrix reads
\begin{equation}
	\rho^{\alpha}_0={1\over Z^{\alpha}_0}
        \exp \left[- \beta(H_{\alpha}-\mu_{\alpha}{N_{\alpha}}) \right]
\end{equation}
where $\beta=1/T$ and $N_{\alpha}=\sum_{k\sigma} a^{\dagger}_{k\sigma{\alpha}}
a_{k\sigma{\alpha}}$ the number of electrons in the lead $\alpha$.
The normalization factor $Z^{\alpha}_0$ is determined by 
$\mbox{tr} \, \rho_0^{\alpha}=1$.
The boson part reads
\begin{equation}
	\rho_0^B={1\over Z^B_0} \exp
	\left[-\beta_B \sum_q \omega_q d_q^{\dagger}d_q\right]\, .
\end{equation}
The temperature of the boson bath, $T_B=1/\beta_B$, may differ in real 
experiments from the electron temperature $T$.

For the initial distribution of the dot we assume that it is diagonal
in the many-body {\it dot} states, $|\chi \rangle$, which include the
strong correlations within the quantum dot but are assumed
to have fixed  occupation numbers,
\begin{equation}
 	\rho^D_0=\sum_{\chi} P^0_{\chi} |\chi \rangle \langle \chi | \, ,
\end{equation}
with $\sum_{\chi} P^0_{\chi}=1$. 
We will see later, that in the stationary limit, i.e., when $t_0$ is shifted 
to minus infinity, all the physical quantities are independent of the choice 
of $P^0_{\chi}$.

In the following, it is convenient to change to the interaction picture with 
respect to $\bar{H}_0$.
This implies $A(t)_{\bar{H}}=\tilde{T}\exp \left(-i\int_t^{t_0} dt' \, 
\bar{H}_T(t')_I \right) A(t)_I \,T\exp \left( -i\int_{t_0}^t dt' \, 
\bar{H}_T(t')_I \right)$ in which $T$ is the time-ordering and $\tilde{T}$ 
denotes the anti-time-ordering operator.
We write the integrals as one contour integral $\int_K dt'\ldots$ over the 
{\it Keldysh contour}. 
It is parameterized by the ``time'' $t'$ which first runs forward from 
$t_0$ to $t$ and then backward from $t$ to $t_0$.
In the diagrammatic language, the Keldysh contour is represented by horizontal
lines running from the left to the right and then back to the left (see 
Fig.~\ref{fig1}).
We find
\begin{equation}\label{general}
	\langle A(t)\rangle = \mbox{tr} \left[ \rho_0 \, T_K \exp \left(-
	i\int_K dt' \bar{H}_T(t')_I \right) A(t)_I \right].
\end{equation}
Here, we have introduced the Keldysh time-ordering operator $T_K$, which 
orders all following operators along the Keldysh contour such that the one 
with the later ``time'' along the Keldysh contour appears at a more left 
position (without any sign change for an exchange of Fermi operators).

In the following we will encounter also higher order correlation functions of 
the type $\langle T_K A_1(t_1)_I A_2(t_2)_I\ldots A_n(t_n)_I \rangle$. 
The final time $t$ of the Keldysh contour is then given by 
$\mbox{max}\{t_1, \ldots , t_n\}$.

For a diagrammatic description we expand the exponential with respect to the 
tunneling Hamiltonian and obtain
\begin{equation}\label{expansion}
	\displaystyle \langle T_K \prod_{i=1}^n A_i(t_i)\rangle = \mbox{tr} \,
	\left[ \rho_0
	\sum_{m=0}^{\infty} (-i)^m
	\int_Kdt_1'\int_Kdt_2'\ldots\int_Kdt_m'\,
        T_K \left\{ {\bar{H}_T(t_1')}_I{\bar{H}_T(t_2')}_I\ldots
	{\bar{H}_T(t_m')}_I \prod_{i=1}^n A_i(t_i)_I \right\} \right]
	\atop 
	\scriptstyle t_1'\, >\, t_2'\, >\, \ldots\, >\, t_m' 
	\qquad\qquad\qquad
\end{equation}
in which the relation $t_1'>t_2'>\ldots >t_m'$ has to be understood with 
respect to the Keldysh contour.
The time-ordering operator $T_K$ acts also on the operators $A_i(t_i)_I$ and 
puts them on the right place between the tunneling Hamiltonians.

The next task is to perform the trace of each term of the expansion. 
We insert the tunnel Hamiltonian (\ref{tunnel}) and notice that the 
Hamiltonian $\bar{H}_0$ is bilinear in the lead electron operators.
For this reason, Wick's theorem holds for these degrees of freedom, i.e., the
lead electron operators are contracted in pairs. 
This includes contractions between pairs of field operators from $\bar{H}_T$ 
as well as contractions to lead electron operators which can be present in the
operators $A_i$.
These contractions are given by equilibrium distribution functions.
For the dot electrons, the situation is different. 
The Coulomb interaction is expressed by a quartic term of dot electron 
operators.
Therefore, Wick's theorem does not hold for this part of the system.
A product of dot electron operators can not be contracted into pairs, but has 
to be treated explicitly. 
The trace over the bosonic part, however, poses no problem. 
Each tunneling term contains an exponential $e^{\pm i\varphi}$ of the bosonic 
operators $\varphi$. 
The trace over the product of such exponentials can be performed easily (see 
Eq.~(\ref{bosontrace})) provided that the operator $A$ also contains only such
exponentials.

\subsection{The rules}

With regard to the applications discussed in section 
\ref{master and probability} and \ref{the current}, we assume here that the 
operators $A_i$ (representing ``external vertices'') depend on the lead and 
boson degrees of freedom only in the form 
\begin{equation}\label{special Ai}
	A_i = A_i\left( 
	\sum_k T^\alpha_k a_{k\sigma\alpha}^\dagger c_\sigma e^{i\varphi}
	\, , \,
	\sum_k {T^\alpha_k}^* c^\dagger_\sigma a_{k\sigma\alpha} e^{-i\varphi}
	\, , \,
	c_\sigma e^{i\varphi}
	\, , \,
	c^\dagger_\sigma e^{-i\varphi}\right)\; .
\end{equation}

Each term of the expansion Eq.~(\ref{expansion}) is visualized by a diagram 
(see Fig.~\ref{fig1}).
There is a forward and a backward propagator symbolized by the upper and lower
horizontal line, running from $t_0$ to $t$ and back from $t$ to $t_0$, 
respectively.
Along this time path, we arrange internal and external vertices according to 
their time-ordering. 
The internal vertices emerge from the insertion of the tunneling Hamiltonian 
(\ref{tunnel}) into the expansion (\ref{expansion}). 
Each of them corresponds to a product of a lead and a dot electron operator 
and a phase factor $e^{\pm i \varphi}$. 
After integrating out the lead degrees of freedom, all vertices (either 
internal or external) containing a lead electron operator are connected in 
pairs by directed tunneling lines (dashed lines) 
$\bar{\gamma}_{\alpha}^K(t,t')$ from $t'$ to $t$, with 
$\bar{\gamma}_{\alpha}^K(t,t')=\bar{\gamma}_{\alpha}^+(t-t')$ for $t<t'$ and 
$\bar{\gamma}_{\alpha}^K(t,t')=\bar{\gamma}_{\alpha}^-(t-t')$ for $t>t'$ with 
respect to the Keldysh contour with
$\bar{\gamma}_{\alpha}^{\pm}(t)=\int {dE}e^{-iE t}
\bar{\gamma}_{\alpha}^{\pm}(E)$.
These tunneling lines represent contractions of lead electron operators.

There are vertices from which a tunneling line leaves (representing 
$a_{k\sigma\alpha}^\dagger (t) c_\sigma (t) e^{i\varphi (t)}$ which removes a
dot electron with spin $\sigma$) and others to which a tunneling line enters 
(visualizing $c_\sigma^\dagger (t) a_{k\sigma\alpha}(t) e^{-i\varphi (t)}$
which adds a dot electron with spin $\sigma$).
Fermi statistics, furthermore, yield a minus sign for each crossing of 
tunneling lines.

In the interaction picture the dot electron operators get exponential factors 
which contain the energies $\epsilon_{\chi}$ of the many-body dot states 
$\chi$ given by $\epsilon_{\chi}|\chi\rangle=\bar{H}_D|\chi\rangle$.
The order of the electron operators may induce furthermore a minus sign due
to Fermi statistics.

The trace over the boson operators gives rise to a factor of the form
\begin{equation}
	C_B(t_1,t_2,\ldots,t_m,t_1',t_2',\ldots,t'_{m'})
	=\langle T_K \left[
	e^{-i\varphi(t_1)}e^{-i\varphi(t_2)}\ldots e^{-i\varphi(t_m)}
	e^{i\varphi(t_1')}e^{i\varphi(t_2')}\ldots e^{i\varphi(t'_{m'})}
	\right] \rangle
\end{equation}
Since $\varphi$ is linear in the boson operators, we get
\begin{equation}\label{bosontrace}
	C_B(t_1,t_2,\ldots,t_m,t_1',t_2',\ldots,t'_{m'})=
	\prod_{i<j}P^K(t_i,t_j)^{-1}
	\prod_{i<j}P^K(t_i',t_j')^{-1}
	\prod_{i,j}P^K(t_i,t_j')\, .
\end{equation}
We write $P^K(t,t')=P^+ (t,t')$ for $t<t'$ and $P^K(t,t')=P^- (t,t')$ for 
$t>t'$ on the Keldysh contour with $P^{\pm}(t)=\int dE e^{-iE t}P^{\pm}(E)$.
In the diagrammatic language, we represent the factors $P^K$ by boson lines
connecting each vertex with each other.

A summary of these rules are given in Appendix~\ref{append rules in time}.

In order to calculate stationary transport properties it is convenient to
change to an energy representation. 
Without loss of generality we assume that the times $t_1,\ldots,t_n$ of the 
correlation function (\ref{expansion}) are ordered on the real axis according 
to $t_n<t_{n-1}<\ldots <t_1=t$.
This may be different to the ordering on the Keldysh contour which depends on 
whether the times lie on the upper or lower branch. 
In the stationary limit we can set $t_0=-\infty$ and $t = t_1 = 0$.

We consider the Laplace transform
\begin{equation}\label{correlation}
	G(E_2,E_3,\ldots,E_n)=(-i)^{n-1}
	\int^0_{-\infty}dt_2\int^{t_2}_{-\infty}dt_3\ldots
	\int^{t_{n-1}}_{-\infty}dt_n\, 
	e^{iE_2 t_2}e^{iE_3 t_3}\ldots e^{iE_n t_n}
	\langle T_K A_1(0) A_2(t_2) \ldots A_n(t_n) \rangle\;.
\end{equation}
We will account for the exponential factors $\exp (iE_i t_i)$ 
($i=2,\ldots,n$) by drawing directed virtual lines from the external vertices 
with time $t_i$ to the last vertex with time $t_1=0$ and assigning the energy 
$E_i$ to this virtual line.

Performing the time integrals we end up with diagrammatic rules in energy 
representation.
These rules are summarized in Appendix~\ref{append rules in energy}.

\subsection{Master equation and stationary probabilities}
\label{master and probability}
 
In this section we will derive a formally exact expression for the central
object of this paper: the quantum-mechanical transition rate 
$\Sigma_{\chi',\chi}(t',t)$ for the reduced system (the dot) to go from a 
state $\chi'$ at time $t'$ to a state $\chi$ at time $t$. 
This rate will serve as an input for a formally exact and time-dependent 
Master equation which, in principle, could be used to calculate all occupation
probabilities of the dot as a function of time for an arbitrary initial state.
Similar Master equations are well-known and successfully applied in connection
with macroscopic quantum coherence phenomena in spin bose models 
\cite{Leg-etal,Wei-Buch,Gri-Sas-Wei}.
The connection to these path-integral approaches will be described in 
Ref.~\onlinecite{HS-Habil}.

A matrix element of the reduced density matrix of the dot at time $t$, 
$P^{\chi_1}_{\chi_2}(t)$, is given by the quantum-statistical expectation 
value of the projector $(|\chi_2 \rangle \langle \chi_1 |) (t)$
\begin{equation}
	P^{\chi_1}_{\chi_2}(t)=\langle(|\chi_2\rangle\langle \chi_1|)(t)
	\rangle \, ,
\end{equation}
i.e., we have to set $n=1$ and $A_1=|\chi_2\rangle\langle\chi_1 |$ in 
Eq.~(\ref{expansion}). 
The reduced density matrix commutes with the particle number on the dot. 
Thus, the operator $|\chi_2\rangle\langle\chi_1 |$ is unaffected by the 
unitary transformation and no tunneling and boson line emerges from this 
external vertex.
The matrix element $P^{\chi_1}_{\chi_2}(t)$ can be expressed by the reduced 
propagator $\Pi^{\chi'_1,\chi_1}_{\chi'_2,\chi_2}(t',t)$ from $\chi'_1$ at 
time $t'$ forward to $\chi_1$ at time $t$ and then from $\chi_2$ at time $t$ 
backward to $\chi'_2$ at time $t'$
\begin{equation}\label{propagation}
	P^{\chi_1}_{\chi_2}(t)=\sum_{\chi'_1,\chi'_2}P^{\chi'_1}_{\chi'_2}(t')
	\Pi^{\chi'_1,\chi_1}_{\chi'_2,\chi_2}(t',t) \;.
\end{equation}
The propagator is the sum of all diagrams with the given states at the ends 
and can be expressed by an irreducible self-energy part 
$\Sigma^{\chi'_1,\chi_1}_{\chi'_2,\chi_2}(t',t)$, defined as the sum of all 
diagrams in which any vertical line cutting through them crosses at least one 
tunneling or boson line. The propagators for the four lines attached to the
self-energy are not included in 
$\Sigma^{\chi'_1,\chi_1}_{\chi'_2,\chi_2}(t',t)$.
We obtain an iteration in the style of a Dyson equation 
(see Fig.~\ref{Fig.dyson1}),
\begin{equation}\label{Dyson}
	\Pi^{\chi'_1,\chi_1}_{\chi'_2,\chi_2}(t',t) = 
	{\Pi^{(0)}}^{\chi_1}_{\chi_2}(t',t) \delta_{\chi_1,\chi_1'} 
	\delta_{\chi_2,\chi_2'} + 
	\sum_{\chi''_1,\chi''_2}\int_{t'}^t dt_2 \int^{t_2}_{t'} dt_1 \, 
	\Pi^{\chi'_1,\chi''_1}_{\chi'_2,\chi''_2}(t',t_1)\, 
	\Sigma^{\chi''_1,\chi_1}_{\chi''_2,\chi_2}(t_1,t_2) \, 
	{\Pi^{(0)}}^{\chi_1}_{\chi_2}(t_2,t) \; ,
\end{equation}
where ${\Pi^{(0)}}^{\chi_1}_{\chi_2}(t',t)=\exp \left[-i(\epsilon_{\chi_1}-
\epsilon_{\chi_2})(t-t')\right]$ is the propagator of the isolated quantum 
dot.
Multiplying this equation with $P^{\chi'_1}_{\chi'_2}(t')$, summing over the 
states $\chi'_1,\chi'_2$ and differentiating with respect to $t$, we obtain 
together with Eq.~(\ref{propagation}) and setting $t'=t_0$
\begin{equation}\label{master-nondiagonal}
	\frac{d}{dt}P^{\chi_1}_{\chi_2}(t)+
	i(\epsilon_{\chi_1}-\epsilon_{\chi_2}) P^{\chi_1}_{\chi_2}(t) = 
	\sum_{\chi'_1,\chi'_2}\int_{t_0}^t dt'\, P^{\chi'_1}_{\chi'_2}(t') 
	\Sigma^{\chi'_1,\chi_1}_{\chi'_2,\chi_2}(t',t)\;.
\end{equation}
This formally exact equation is the most general kinetic equation for the
reduced density matrix of the dot. 
No assumption is necessary for the initial state and the integral kernel 
$\Sigma$ on the r.h.s. shows that memory effects are fully taken into account.

The equation simplifies considerably if we assume that the initial density
matrix is diagonal. 
In the general case, this does not imply that the reduced density matrix stays
diagonal for all times. 
However, for the special case of the Anderson model considered here, spin 
conservation implies that the reduced density matrix will be diagonal for all 
times $t>t_0$. 
Hence, we consider $\Sigma_{\chi',\chi}\equiv\Sigma^{\chi',\chi}_{\chi',\chi}$
and obtain from (\ref{master-nondiagonal})
\begin{equation}\label{master}
	\frac{d}{dt}P_{\chi}(t) = \sum_{\chi'}\int_{t_0}^t dt'\,
	P_{\chi'}(t') \Sigma_{\chi',\chi}(t',t)\;,
\end{equation}
where $P_{\chi}(t)\equiv P^{\chi}_{\chi}(t)$ denotes the probability to be in 
the state $\chi$ at time $t$. 
For a time-translational invariant system, the time-dependent rates 
$\Sigma_{\chi',\chi}(t',t)$ depend only on the time difference 
$\Sigma_{\chi',\chi}(t'-t)$.
Performing the Laplace transform of Eq.~(\ref{master}) one can then study the 
time evolution of arbitrary initial probability distributions into the 
stationary state.

By attaching the rightmost vertex of each diagram $\Sigma$ to the upper and 
lower propagator the minus sign for each vertex on the backward propagator 
yields $\sum_{\chi}\Sigma_{\chi',\chi}(t',t)=0$, which allows us to rewrite 
Eq.~(\ref{master}) in the form
\begin{equation}\label{time master}
	\frac{d}{dt}P_{\chi}(t) = \sum_{\chi'\ne\chi}\int_{t_0}^t dt'\,
	\left[ P_{\chi'}(t') \Sigma_{\chi',\chi}(t',t)
 	-P_{\chi}(t') \Sigma_{\chi,\chi'}(t',t)\right]\;,
\end{equation}
We obtain the structure of a master equation with transition rates given by 
$\Sigma_{\chi',\chi}(t',t)$. 

The stationary distribution is given by
\begin{equation}
	P_{\chi}^{st}=\lim_{t\rightarrow\infty}P_{\chi}(t)
	=\lim_{t_0\rightarrow -\infty}P_{\chi}(0)
\end{equation}
and is {\it not} the equilibrium one if the electrochemical potentials of the 
leads are different. 
From (\ref{master}) and (\ref{time master}) we obtain
\begin{equation}\label{stationary master}
	0=\sum_{\chi'}P^{st}_{\chi'} {\Sigma}_{\chi',\chi}  = 
	\sum_{\chi'\ne\chi} [ P^{st}_{\chi'} \Sigma_{\chi',\chi} 
             - P^{st}_{\chi} \Sigma_{\chi,\chi'}] \; ,
\end{equation}
where 
\begin{equation}\label{self-energy}
	\Sigma_{\chi',\chi}=i\int_{-\infty}^0 dt'\,\Sigma_{\chi',\chi}(t',0)
\end{equation}
can be calculated directly by using our diagrammatic rules in energy space.
The prefactor $i$ in (\ref{self-energy}) together with the $m-1$ 
time integrations over the internal vertices in $\Sigma$ gives a factor $i^m$ 
in (\ref{integral}) which cancels the factor $(-i)^m$ from rule 5.

The self-energy part $\Sigma_{\chi',\chi}$ is purely imaginary. 
This can be seen by changing the vertical positions of all vertices on the 
Keldysh contour (without changing their horizontal position) and reversing the
direction of all tunneling and boson lines. 
Consequently, only the energy differences $\Delta E_j$ from rule 2' will 
change sign. 
Since the number of vertices is even (all tunneling vertices are coupled in 
pairs by tunneling lines), there is no sign change due to rule 5' and the 
number of resolvents is odd. 
Thus, the whole diagram has been changed to its conjugate complex up to a 
sign.

\subsection{The tunneling current}
\label{the current}

The tunneling current flowing into reservoir $\alpha$ is defined by 
$I_{\alpha}(t)=e{d\over dt}{\langle N_{\alpha} (t)\rangle}
=ie \langle [\bar{H},N_{\alpha}](t)\rangle$, which is equivalent to 
\begin{equation}\label{current}
	I_\alpha(t)=-ie \sum_{k\sigma}\left\{
 	T^{\alpha}_{k}{\langle(a^\dagger_{k\sigma\alpha}
	c_{\sigma}e^{i\varphi})(t)\rangle}\,-\,
 	{T^{\alpha}_{k}}^*{\langle(c^\dagger_{\sigma}
 	a_{k\sigma\alpha}e^{-i\varphi})(t)\rangle}\right\}\, .
\end{equation}

The tunneling current is an expectation value of a product of a dot, boson and
reservoir electron operator (see Fig.~\ref{currentdot}). 
We obtain
\begin{equation}\label{current-sigma}
	I_\alpha (t)=e\sum_{\chi,\chi'}\int_{t_0}^t dt' P_{\chi'}(t')
	\Sigma^{\alpha+}_{\chi',\chi}(t',t)
	= -e\sum_{\chi,\chi'}\int_{t_0}^t dt' P_{\chi'}(t')
	\Sigma^{\alpha-}_{\chi',\chi}(t',t)
\end{equation}
where the partial self-energies $\Sigma^{\alpha \pm}_{\chi',\chi}(t',t)$ 
are parts of the total self-energy
\begin{equation}\label{rate-partial rate}
	\Sigma_{\chi',\chi}(t',t)=
	\sum_{\alpha} \left\{ \Sigma^{\alpha+}_{\chi',\chi}(t',t) +
	\Sigma^{\alpha-}_{\chi',\chi}(t',t) \right\} \, .
\end{equation}
They describe processes in which the rightmost tunneling line corresponds to 
reservoir $\alpha$ and is an outgoing (incoming) line if the rightmost vertex 
lies on the upper propagator or an incoming (outgoing) line if the rightmost 
vertex lies on the lower propagator. 
Their physical meaning is displayed by the current formula 
(\ref{current-sigma}) which shows that they give the total contribution to the
current rate. 
We can relate them to an intuitively more physical object, namely the rate 
$\Sigma_{\chi',\chi}^{\alpha p}(t',t)$, $p=0,\pm 1,\pm 2,\ldots$, which 
describes the transition rate where $p$ particles are transferred to reservoir
$\alpha$. 
Within our graphical language $\Sigma_{\chi',\chi}^{\alpha p}(t',t)$ is given 
by all diagrams where the number of tunneling lines with reservoir index 
$\alpha$ running from the forward to the backward propagator minus the number 
of tunneling lines with reservoir index $\alpha$ running from the backward to 
the forward propagator is given by $p$. 
We obtain
\begin{equation}\label{rate relation}
	\sum_{\chi}\Sigma^{\alpha\pm}_{\chi',\chi}(t',t)=\pm
	\sum_{\chi}\sum_p p\,\Sigma^{\alpha p}_{\chi',\chi}(t',t).
\end{equation}
This relation together with current conservation is proven in 
Appendix~\ref{append proof}.
The factor $p$ shows clearly that $\Sigma^{\alpha\pm}$ describes the 
contribution to the current rate. 
In contrast to lowest order processes, i.e., the golden rule rate, where $p$ 
can only take the values $\pm 1$, $p$ can be arbitrary for higher order 
processes.
Nevertheless, Eq.~(\ref{rate-partial rate}) shows that the current rate can be
calculated as a partial selection of diagrams already contained in the total 
transition rate $\Sigma_{\chi',\chi}$.

We emphasize that the current formula (\ref{current-sigma}) together with the 
Master equation (\ref{master}) constitutes a complete theory to describe 
time-dependent phenomena starting from an arbitrary diagonal initial state. 
The original problem has now been shifted to the evaluation of the various 
self-energy diagrams which correspond to transition and current rates. 
The self-energies are defined by a set of irreducible diagrams and thus their 
corresponding perturbation expansion in the number of tunneling lines is a 
well-defined series and contains no divergent time-integrals.

For time-translational invariant systems the current rates 
$\Sigma^{\alpha\pm}_{\chi',\chi}(t',t)$ depend only on the time difference
$t'-t$. 
To calculate the stationary current we define in analogy to 
(\ref{self-energy})
\begin{equation}\label{current-rates}
	\Sigma^{\alpha\pm}_{\chi',\chi}=
	i\int_{-\infty}^0 dt' \Sigma^{\alpha\pm}_{\chi',\chi}(t',0)
\end{equation}
which again can be calculated directly with our diagrammatic rules in energy 
space. 
The stationary current is then given by
\begin{equation}\label{stationary current}
	I_\alpha^{st}=-ie\sum_{\chi,\chi'} P^{st}_{\chi'} 
	\Sigma_{\chi',\chi}^{\alpha +}
	=ie\sum_{\chi,\chi'} P^{st}_{\chi'} \Sigma_{\chi',\chi}^{\alpha -}\, .
\end{equation}

\subsection{Green's functions}

After the unitary transformation the Green's functions of the dot electrons 
read
\begin{eqnarray}
	G^>_{\sigma}(t,t')&=&-i \langle (c_{\sigma}e^{i\varphi})(t)
	(c^{\dagger}_{\sigma}e^{-i\varphi})(t')\rangle
	\\
	G^<_{\sigma}(t,t')&=&i \langle (c^{\dagger}_{\sigma}
	e^{-i\varphi})(t') (c_{\sigma}e^{i\varphi})(t) \rangle \, .
\end{eqnarray}
Here $G^>_{\sigma}$ and $G^<_{\sigma}$ are independent quantities since we do 
not assume equilibrium.
For time-translational invariant systems, the Green's functions depend only on
the time difference $G(t,t')=G(t-t')$.
The Fourier transform $G(E)=\int dt e^{i E t} G(t)$ can be written in the form
\begin{eqnarray}
	G^>_{\sigma}(E)&=&2i \, \mbox{Im} \, (-i) \int_{-\infty}^0 dt 
	e^{-iE t}\langle 
	T_K(c_{\sigma}e^{i\varphi})(0)(c^{\dagger}_{\sigma}e^{-i\varphi})(t^+)
	\rangle 
	\\
	G^<_{\sigma}(E)&=&-2i \, \mbox{Im} \, (-i) \int_{-\infty}^0 dt 
	e^{-iE t}\langle 
	T_K(c_{\sigma}e^{i\varphi})(0)(c^{\dagger}_{\sigma}e^{-i\varphi})(t^-)
	\rangle \, ,
\end{eqnarray}
where $t^{\pm}$ means that the the time $t$ lies on the upper (lower) branch 
of the Keldysh contour. 
Note that the time ordering is defined here by a pure ordering along the 
Keldysh contour without any sign change if we interchange fermion operators. 
The integrals can be calculated like Eq.~(\ref{correlation}) whereby instead 
of assigning the energy $-E$ to the virtual line connecting the external 
vertices one can change the direction of the line and assign the energy $E$.

In order to relate the current to the Green's functions of the dot we consider
the first diagram on the r.h.s. of Fig.~\ref{currentdot} (the second one is 
just the conjugate complex). 
The external vertex can be either contracted by a tunneling line to the upper 
or lower propagator and we recover immediately the structure of the Green's 
functions $G^>$ and $G^<$, respectively (see Fig.~\ref{currentcorr}).
We recover for the stationary current the relation
\cite{Car-etal,Her-Dav-Wil,Mei-Win-Lee}
\begin{equation}
	I^{st}_\alpha=-ie \sum_\sigma \int dE \left\{ 
	\gamma^+_\alpha(E)G_\sigma^>(E)+
	\gamma^-_\alpha(E)G_\sigma^<(E)\right\}.
\end{equation}

In the case that the couplings to the leads have the same energy dependence,
$\Gamma_{\alpha}(E) / \Gamma_{\alpha'}(E)=\lambda_{\alpha,\alpha'}$, this can 
be written in the form (which was already derived in 
Ref.~\onlinecite{Mei-Win-Lee}),
\begin{equation}\label{strom}
	I^{st}_{\alpha}=e\sum_{\alpha'}\sum_{\sigma}\int dE \,
	{\Gamma_{\alpha}(E) \Gamma_{\alpha'}(E) \over 
	\sum_{\alpha''}\Gamma_{\alpha''}(E)} \rho_{\sigma} (E) 
	[f^+_{\alpha'} (E)-f^+_{\alpha} (E)] \;.
\end{equation} 
Here, we used the relation between the Green's functions 
$G^<_{\sigma}, G^>_{\sigma}$ and spectral density 
$\rho_{\sigma} \equiv {G^<_{\sigma} - G^>_{\sigma}\over 2\pi i}$.

\section{Results}
\label{results}

What we have done so far is to derive a diagrammatic language which allows
a systematic description of transport processes.
We have, furthermore, shown how the physical quantities of interest, the 
stationary probability distribution and the current, can be obtained if we
know the value of special diagrams.
In this section, now, we will explicitly calculate the value of the 
corresponding diagrams.

We consider here the case of strong Coulomb repulsion $U$, i.e., we restrict 
ourselves to the states with $N=0,1$.
Diagrams in which a higher occupancy occurs do not contribute since they have 
resolvents of the order $1/U$.

In the following, the index $\sigma$ labels the singly occupied state with
spin $\sigma=1,\ldots M$.
The label $\chi$ additionally allows an empty dot, $\chi=0,1,\ldots M$.

In general, we can not sum up {\it all} possible diagrams.
Therefore, we have to find a systematic criterion which diagrams should be
retained and summed.

\begin{figure}
\centerline{\psfig{figure=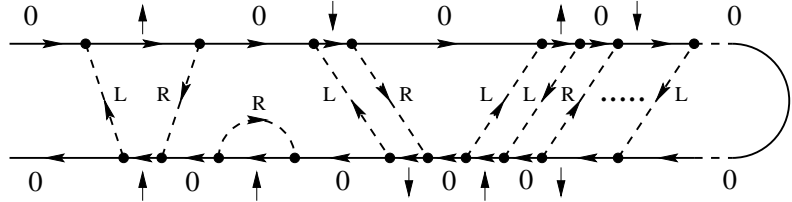,height=2.7cm}}
\vspace{0.3cm}
\caption{A diagram showing various tunneling processes: sequential tunneling 
	in the left and right junctions, a term preserving the norm, a 
	cotunneling process, and resonant tunneling.}
\label{fig1}
\end{figure}

\begin{figure}
\centerline{\psfig{figure=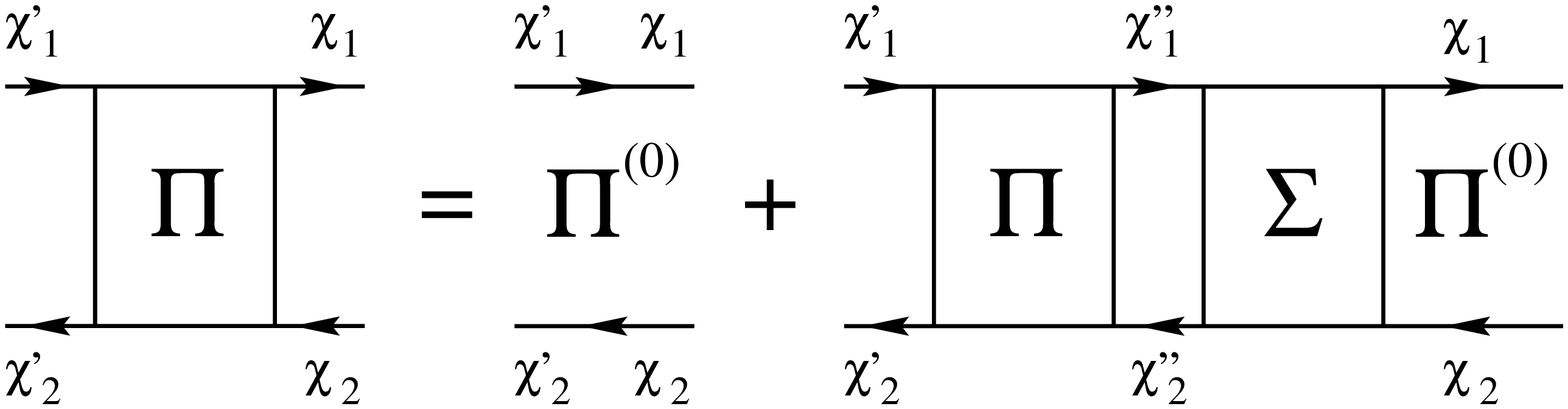,height=3cm}}
\caption{The iteration of processes for the propagator $\Pi$.}
\label{Fig.dyson1}
\end{figure}

\begin{figure}
\centerline{\psfig{figure=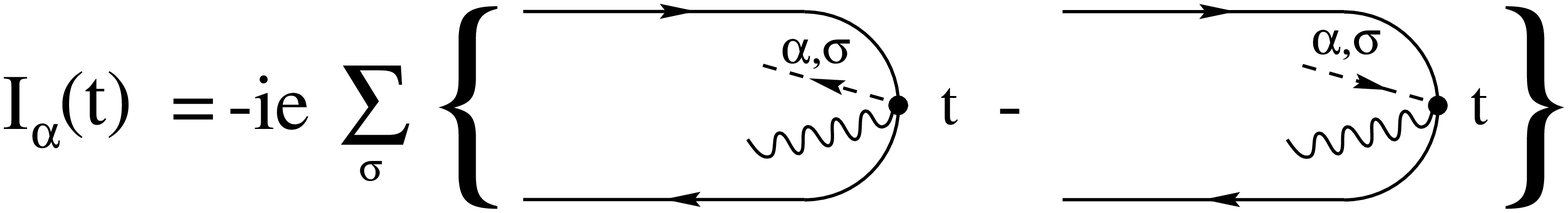,height=1.5cm}}
\caption{Graphical representation of the current $I_\alpha$ through lead 
	$\alpha$.
	Internal vertices are not indicated.}
\label{currentdot}
\end{figure}

\begin{figure}
\centerline{\psfig{figure=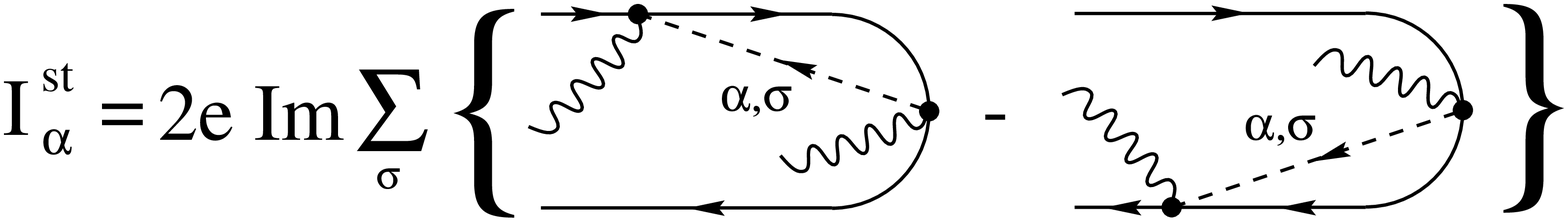,height=1.5cm}}
\caption{Graphical representation of the relation between the current and
	the correlation functions.
	Here, the line connecting the external vertices is a real one.
	Internal vertices are not indicated.}
\label{currentcorr}
\end{figure}

The simplest approximation is to neglect all diagrams where two or more 
tunneling lines overlap in time (see the leftmost diagram parts in
Fig.~\ref{fig1}).
This means we include those processes which are also described by the master 
equation with rates obtained in lowest order perturbation theory (sequential 
tunneling), which is a good description at high temperature, $\Gamma \ll T$.

In situations when sequential tunneling is suppressed by Coulomb blockade,
the lowest order contribution to the current arises due to cotunneling.
The rates for a process in which an electron enters the dot from the left lead
and leaves to right one is described by diagrams with two overlapping lines
(see the diagram part in the middle of Fig.~\ref{fig1}).

At lower temperature the perturbative approach is not sufficient.
Higher-order processes become important.
In generalization to cotunneling we have to take into account irreducible 
diagrams with an arbitrary number of correlated tunneling processes, i.e.,
we include resonant tunneling.

Similar to the case of metallic islands \cite{HS-Sch,Koe-HS-Sch1} we proceed 
in a conserving approximation, taking into account non-diagonal matrix 
elements of the total density matrix up to the difference of one electron-hole
pair excitation in the leads.  
The graphical representation of this constriction is that only diagrams in 
which any vertical line will cut at most two tunneling lines are taken into 
account.

We give two arguments why this class of diagrams is the most important one.
Firstly, since we treat the leads as large equilibrium reservoirs there should
be a tendency of the system to stay close to diagonal states.
Secondly, our approximation contains the exact solution for the noninteracting
limit $U=0$: if there is no electron-electron interaction in the dot, 
electrons with different spin do not influence each other, so that this limit 
is described within our model by choosing $M=1$. 
In this case, the selected diagrams are the only contributing ones.
The sum of all other, more complicated, diagrams is zero.

Furthermore, we include only boson lines between vertices which are already 
connected by tunneling lines, i.e.,
\begin{equation}
	C_B(t_1,t_2,\ldots,t_m,t_1',t_2',\ldots,t_m') \approx
	\prod_{i=1}^m P^K(t_i,t_i')\, .
\end{equation}
where the pairs $t_i,t'_i$ are already coupled by tunneling lines running
from $t'_i$ to $t_i$.
This amounts to a dressing of the tunneling lines $\bar{\gamma} \rightarrow 
\gamma$.
This approximation, while neglecting many diagrams, describes well the 
spectral density of the dot at resonance points.
The reason is that position and value of the peaks of the spectral density are
determined by a self-energy $\sigma$ (see Eq.~(\ref{sigma2})) which is 
calculated here in lowest order perturbation theory in $\Gamma$ including the 
bosons. 
Higher orders are small for high tunnel barriers. 

\begin{figure}
\centerline{\psfig{figure=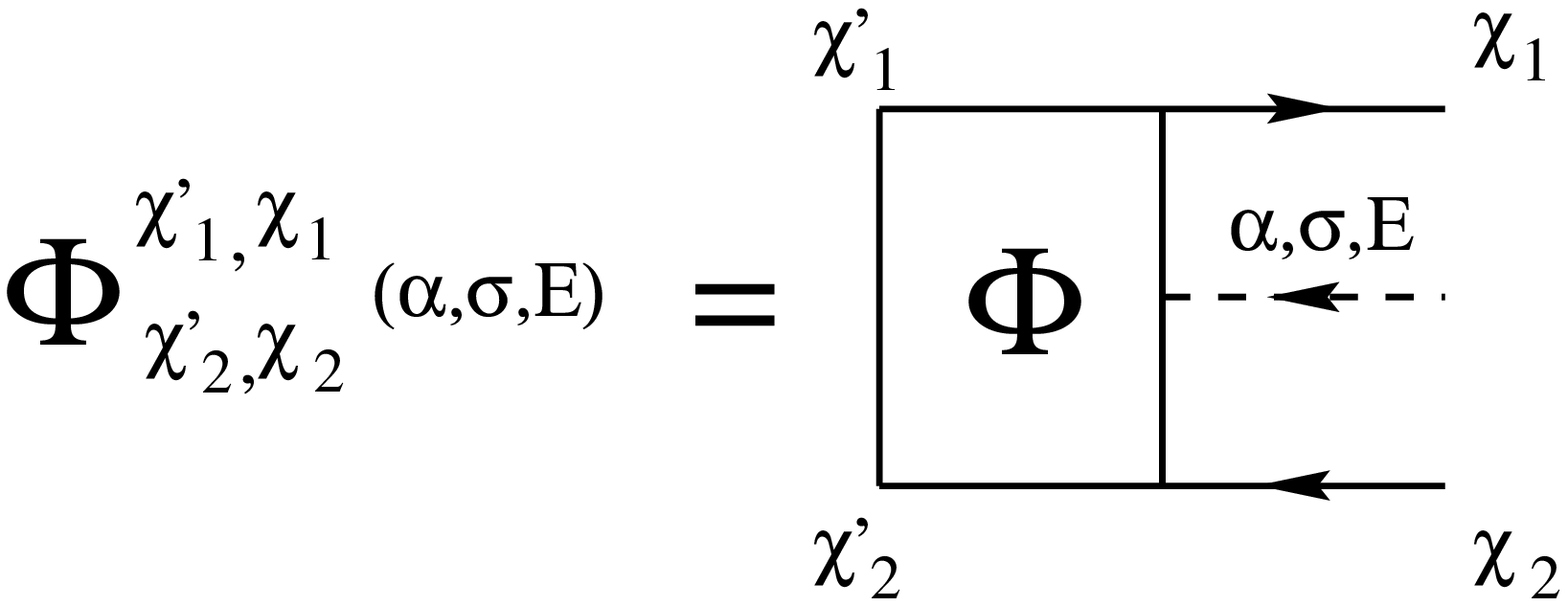,height=3cm}}
\caption{Definition of $\phi$. 
	It denotes a part of a diagram with an open tunneling line entering 
	from the right.}
\label{Fig.phi}
\end{figure}

\begin{figure}
\centerline{\psfig{figure=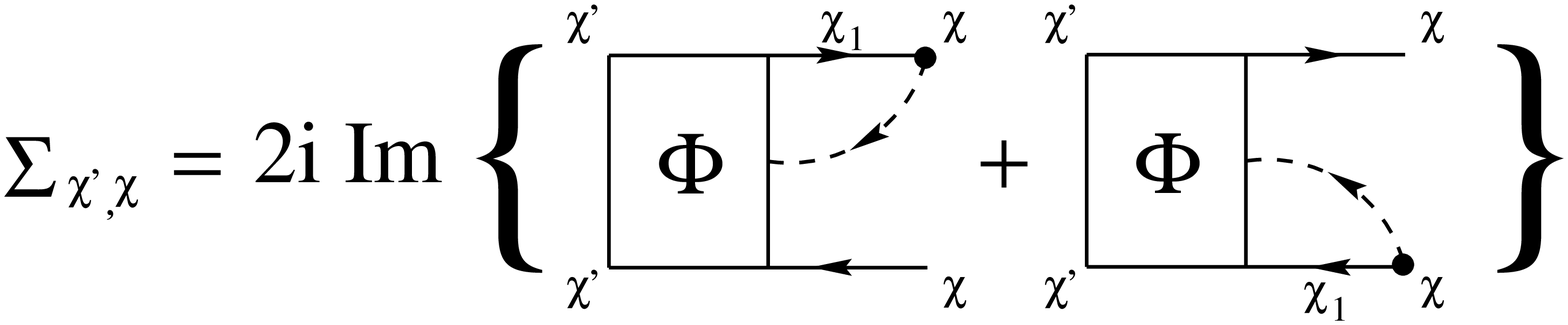,height=2.5cm}}
\caption{The irreducible self energy is obtained by attaching the open 
	tunneling line of $\phi$ and $\phi^*$ to the upper and lower 
	propagator.}
\label{sigmaphi}
\end{figure}

First, we relate the rate $\Sigma_{\chi',\chi}$ to an irreducible diagram 
labeled by $\phi^{\chi'_1,\chi_1}_{\chi'_2,\chi_2}(\alpha,\sigma,E)$ 
(see Fig.~\ref{Fig.phi}).
It has an open tunneling line associated with tunneling of an electron with 
spin $\sigma$ in the junction $\alpha$ carrying the energy $E$. 
The line is directed from the right to the left and its value together with
the corresponding resolvent is included in $\phi$.
The self-energy is then constructed by attaching the open tunneling line of 
these diagrams to the upper and lower propagator (see Fig.~\ref{sigmaphi})
with the result
\begin{eqnarray}\label{partial rates}
	\Sigma_{\chi',\chi} &=& 2i \, \mbox{Im} \int dE \sum_{\sigma,\alpha}
	\sum_{\chi_1} \left\{ \langle \chi | c_\sigma | \chi_1 \rangle 
	\phi^{\chi',\chi_1}_{\chi',\chi} (\alpha,\sigma,E) -
	\langle \chi_1 | c_\sigma | \chi \rangle
	\phi^{\chi',\chi}_{\chi',\chi_1} (\alpha,\sigma,E) \right\}
	\nonumber \\
	&=& \sum_{\alpha} \left\{ \Sigma^{\alpha+}_{\chi',\chi} +
	\Sigma^{\alpha-}_{\chi',\chi} \right\} \, .
\end{eqnarray}
where the current rates $\Sigma^{\alpha\pm}_{\chi',\chi}$ correspond to the 
first and second term, respectively.
Again we have made use of the fact that a diagram becomes the conjugate 
complex if we change the vertical position of all vertices and the direction 
of all tunneling and boson lines.
As pointed out in Appendix~\ref{append proof}, any approximation for $\phi$ 
will lead to a current conserving theory.

\begin{figure}
\centerline{\psfig{figure=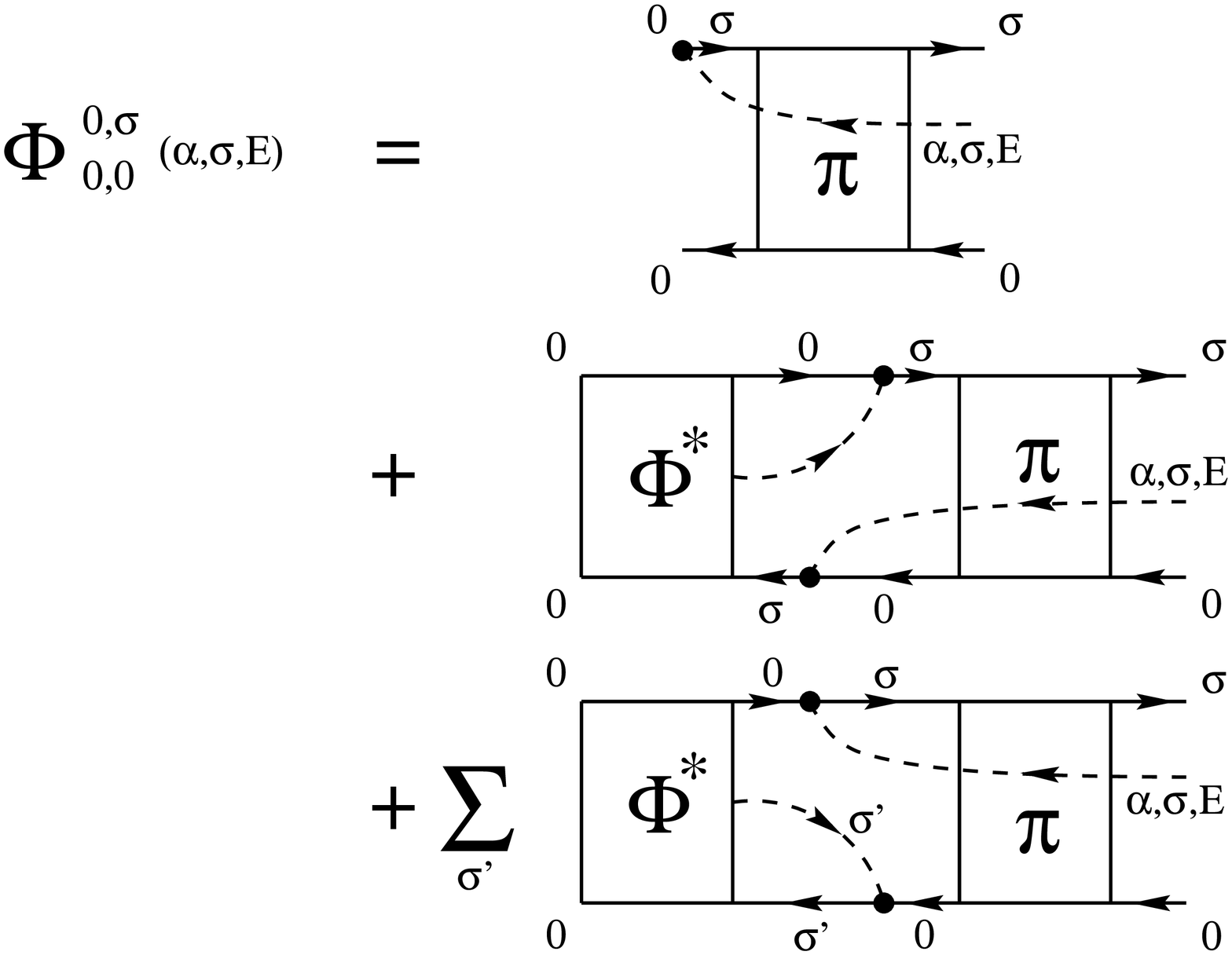,height=7cm}}
\caption{Graphical representation of the self-consistent equation for $\phi$
	beginning with an empty dot state.}
\label{phidot1}
\end{figure}

\begin{figure}
\centerline{\psfig{figure=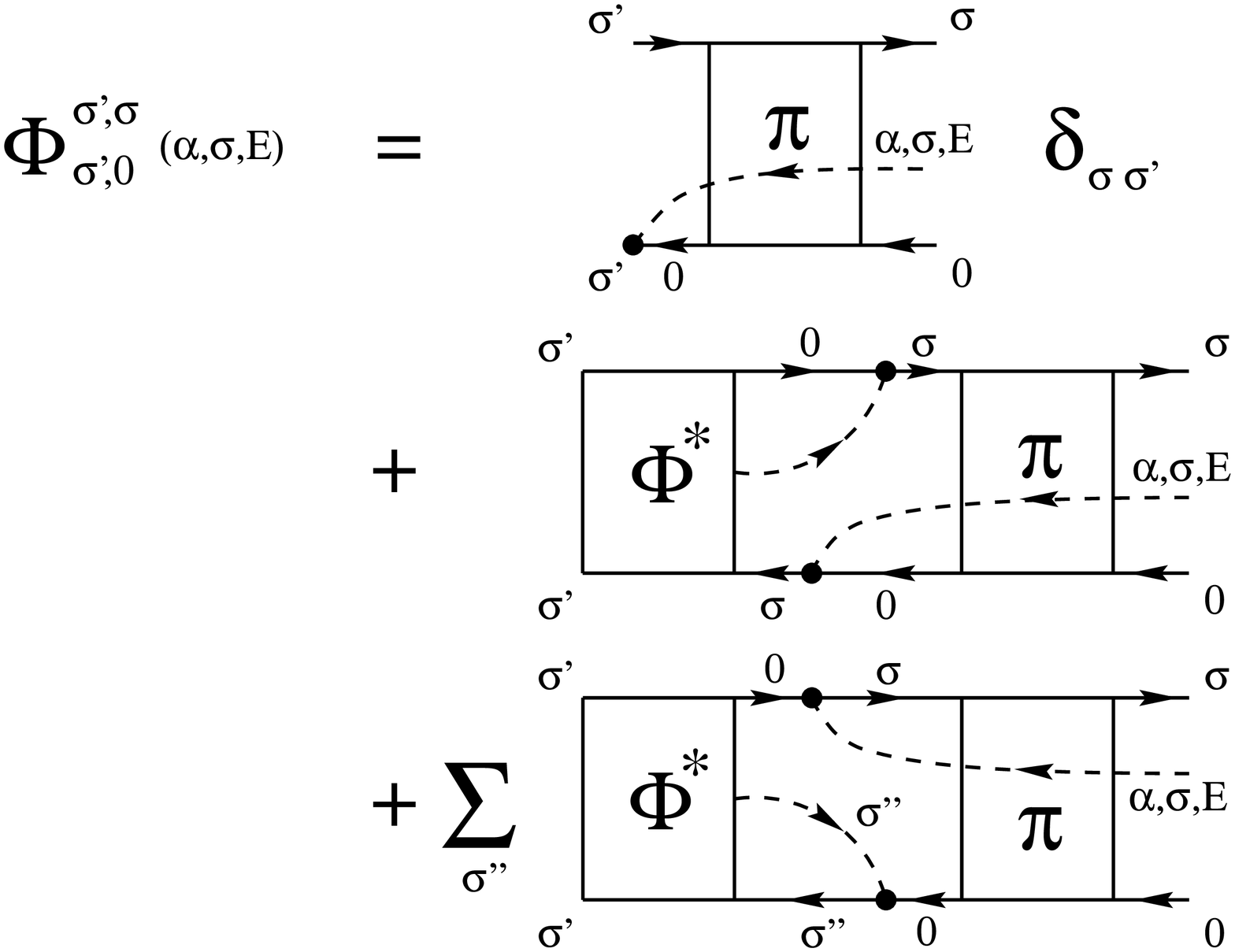,height=7cm}}
\caption{Graphical representation of the self-consistent equation for $\phi$
	beginning with an occupied dot state.}
\label{phidot2}
\end{figure}

\begin{figure}
\centerline{\psfig{figure=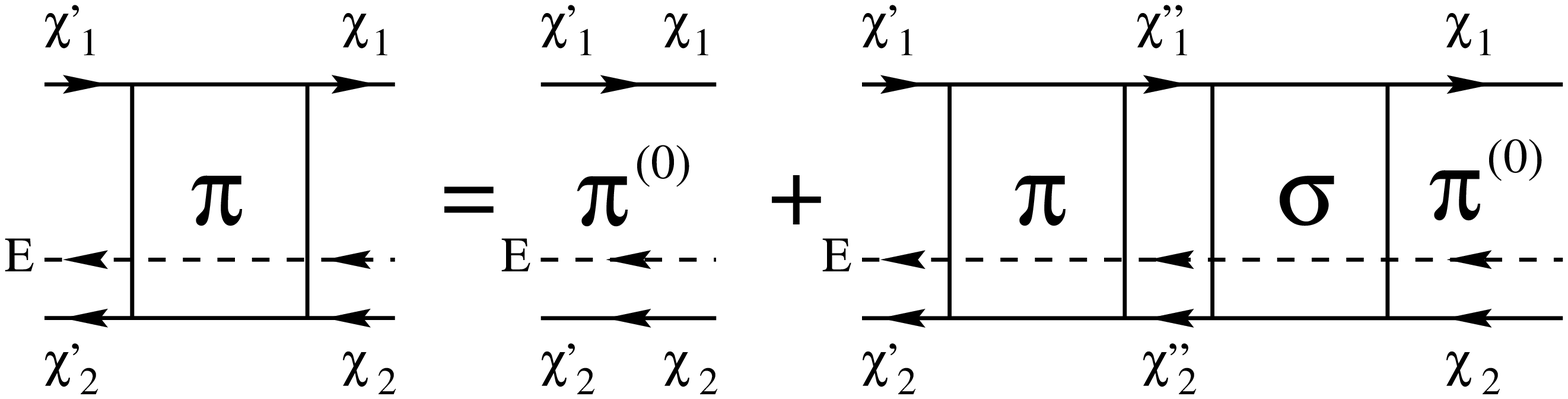,height=3cm}}
\caption{The iteration of processes for the propagator $\pi$ with a tunneling
	line running in parallel from the right to the left.}
\label{Fig.dyson2}
\end{figure}

We construct the diagram $\phi$ by iteration (see Figs.~\ref{phidot1} and 
\ref{phidot2}).
To do so, we need the diagram $\pi (E)$ which is the propagator $\pi (E)$ 
while a tunneling line with energy $E$ is running in parallel from the right 
to the left.
This diagram can also be expressed as an iteration in the style of a Dyson 
equation (see Fig.~\ref{Fig.dyson2})
\begin{equation}\label{pi-dyson}
	\pi^{\chi'_1,\chi_1}_{\chi'_2,\chi_2} (E)= 
	{\pi^{(0)}}^{\chi'_1}_{\chi'_2} (E) \delta_{\chi'_1,\chi_1} 
	\delta_{\chi'_2,\chi_2} + \sum_{\chi''_1,\chi''_2}
	\pi^{\chi'_1,\chi''_1}_{\chi'_2,\chi''_2}(E)\, 
	\sigma^{\chi''_1,\chi_1}_{\chi''_2,\chi_2}(E) \, 
	{\pi^{(0)}}^{\chi_1}_{\chi_2} (E)\; .
\label{dyson2}
\end{equation}

In analogy to $\Sigma$, the self-energy $\sigma(E)$ denotes the sum of all 
irreducible diagrams with a tunneling line going backward in time.
Here, the free propagator in energy space is given by
\begin{equation}
	{\pi^{(0)}}^{\chi_1}_{\chi_2}(E)= 
	{1\over E - (\epsilon_{\chi_1} - \epsilon_{\chi_2})+i0^+}
\end{equation}
Hence, we can solve Eq.~(\ref{pi-dyson}) and find in matrix notation the 
general relation
\begin{equation}\label{pi}
	\pi(E)=[[\pi^{(0)}(E)]^{-1} - \sigma(E)]^{-1} \, .
\end{equation}

\begin{figure}
\centerline{\psfig{figure=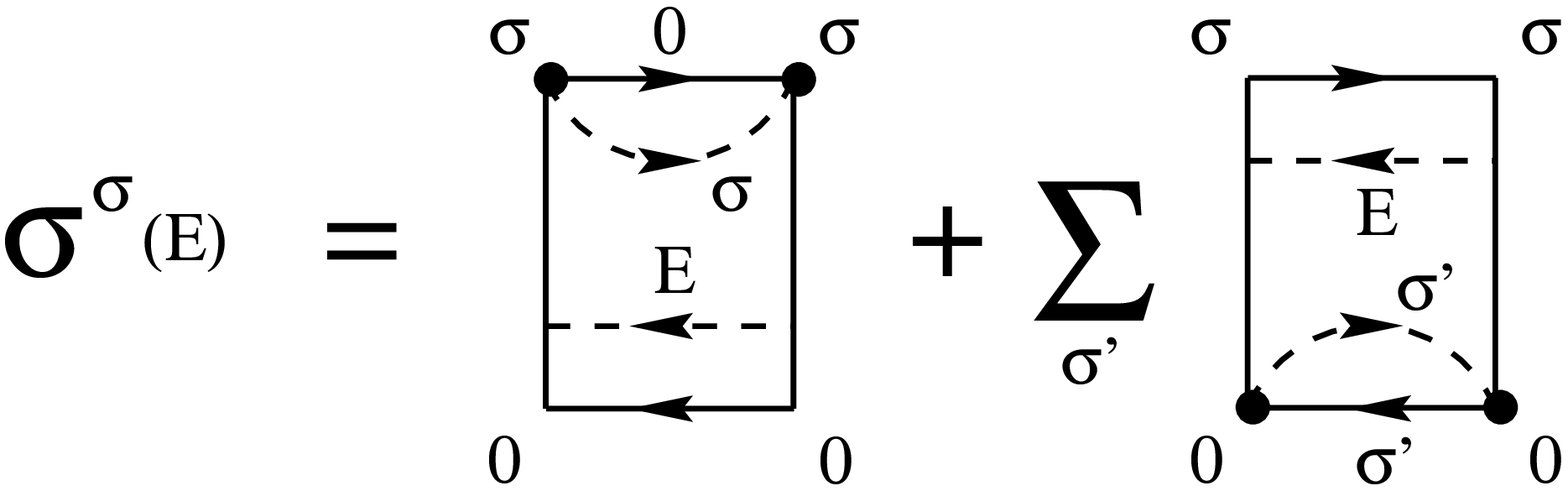,height=2.5cm}}
\caption{In our approximation, the diagram for the irreducible self-energy 
	$\sigma^\sigma(E)$ contains one tunneling line in addition to the 
	backward running line.}
\label{somegadot}
\end{figure}

Because of the restriction to two charge states, only the matrix elements 
$\pi^\sigma(E)\equiv\pi^{\sigma,\sigma}_{0,0}(E)$ of $\pi(E)$ and
$\sigma^\sigma(E)\equiv\sigma^{\sigma,\sigma}_{0,0}(E)$ of 
$\sigma(E)$ are involved, and we deduce from Eq.~(\ref{pi})  
\begin{equation}\label{pi2}
	\pi^\sigma(E)=
	{1\over E-\epsilon_\sigma-\sigma^\sigma(E)} \, . 
\end{equation}
Since at most two tunneling lines are allowed at once, the irreducible 
self-energy $\sigma^\sigma(E)$ consists of only one tunneling line.
We calculate all contributions, which are depicted in Fig.~\ref{somegadot}, 
and get
\begin{equation}\label{sigma2}
	\sigma^\sigma(E)=\int dE'{\gamma^-(E')\over
	E-E'+i0^+} +
	\sum_{\sigma'}\int dE'{\gamma^+(E')\over
	E-E'+\epsilon_{\sigma'}-\epsilon_\sigma+i0^+}\, . 
\end{equation}
In the spin degenerate case, this is exactly the relation (\ref{sigma}) which
we found from intuitive arguments.

According to our rules, Figs.~\ref{phidot1} and \ref{phidot2} lead to the 
self-consistent equation for the diagram $\phi(\alpha,\sigma,E)$
\begin{eqnarray}
	\phi_{0,0}^{0,\sigma}(\alpha,\sigma,E)=
	\pi^{\sigma}(E)\left[\gamma^+_\alpha(E)
	- \gamma^-_\alpha(E)\sum_{\alpha'}\int dE'
	{1\over E-E'+i0^+}
	\,{\phi^*}_{0,0}^{0,\sigma}(\alpha',\sigma,E')\right.
	\nonumber \\
	- \left. \gamma^+_\alpha(E)\sum_{\sigma'}\sum_{\alpha'} 
	\int dE' 
	{1\over E-E'+\epsilon_{\sigma'}-\epsilon_\sigma+i0^+}
	\,{\phi^*}_{0,0}^{0,\sigma'}(\alpha',\sigma',E')\right]
\end{eqnarray}
and
\begin{eqnarray}
	\phi_{\sigma',0}^{\sigma',\sigma}(\alpha,\sigma,E)=
	\pi^{\sigma}(E)
	\left[-\gamma^-_\alpha(E)\delta_{\sigma \sigma'}
	- \gamma^-_\alpha(E)\sum_{\alpha'}\int dE'
	{1\over E-E'+i0^+}
	\,{\phi^*}_{\sigma',0}^{\sigma',\sigma}(\alpha',\sigma,E')\right.
	\!\!\!\!\!\!\!\!\! \nonumber \\
	- \left. \gamma^+_\alpha(E)\sum_{\sigma''}\sum_{\alpha'} 
	\int dE'
	{1\over E-E'+\epsilon_{\sigma'}-\epsilon_\sigma+i0^+}\,
	{\phi^*}_{\sigma',0}^{\sigma',\sigma''}(\alpha',\sigma'',E')
	\right]\,\,
\end{eqnarray}

The stationary probabilities and the current are derived from 
Eqs.~(\ref{stationary master}) and (\ref{stationary current}). 
To calculate the rates we specify Eq.~(\ref{partial rates}) and obtain 
\begin{eqnarray}\label{rates-special}
	\Sigma_{\chi',0}^{\alpha+}&=&2i\,\mbox{Im} \sum_{\sigma} \int dE \,
	\phi^{\chi',\sigma}_{\chi',0}(\alpha,\sigma,E) \, , \\
	\Sigma_{\chi',\sigma}^{\alpha-}&=&-2i\,\mbox{Im} \int dE \,
	\phi^{\chi',\sigma}_{\chi',0}(\alpha,\sigma,E) \, ,
\end{eqnarray}
whereas all other rates are zero.

\begin{figure}
\centerline{\psfig{figure=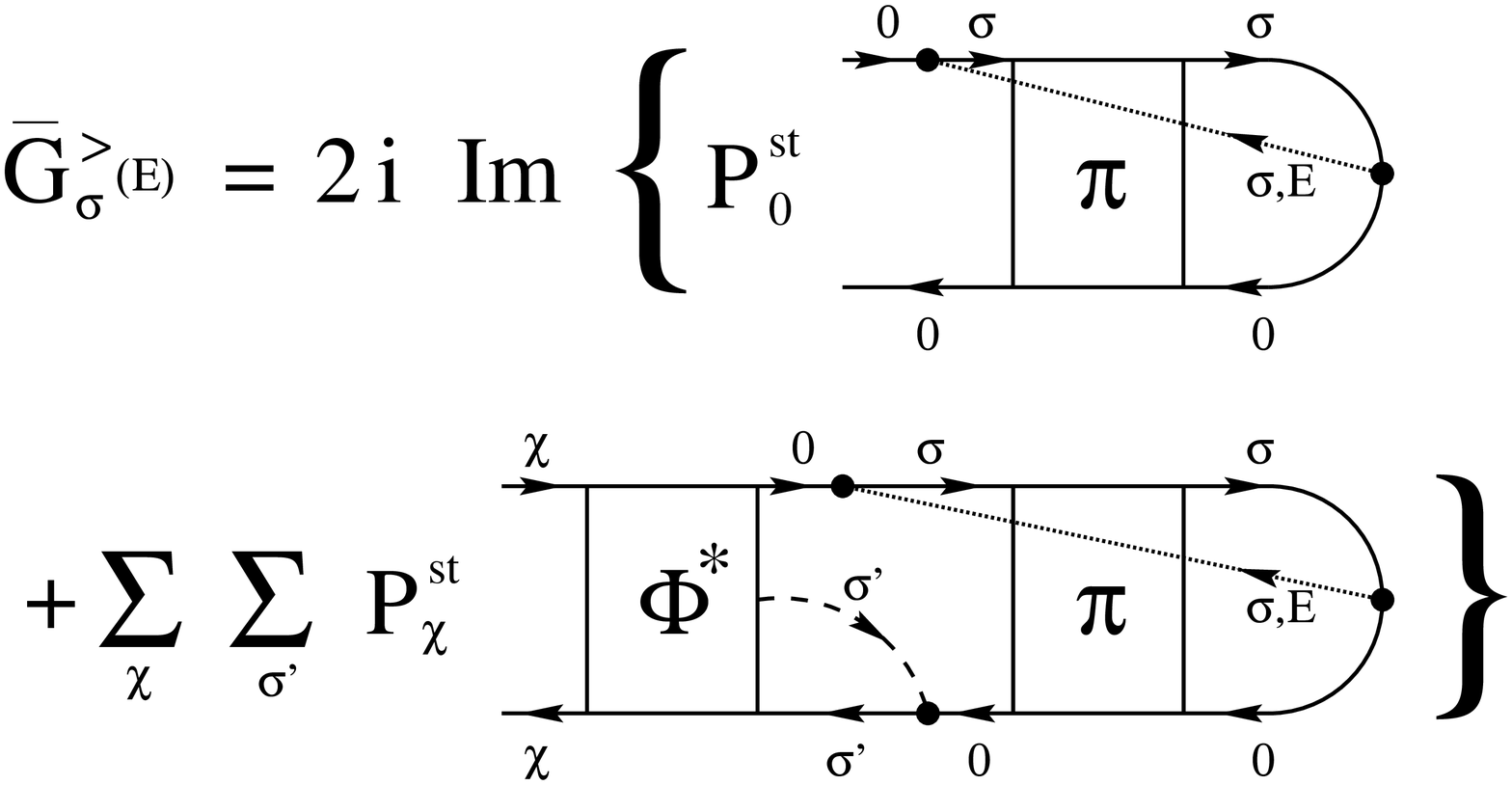,height=5cm}}
\caption{Graphical representation of $\bar{G}_\sigma^>(E)$.}
\label{Cdotgr}
\end{figure}

\begin{figure}
\centerline{\psfig{figure=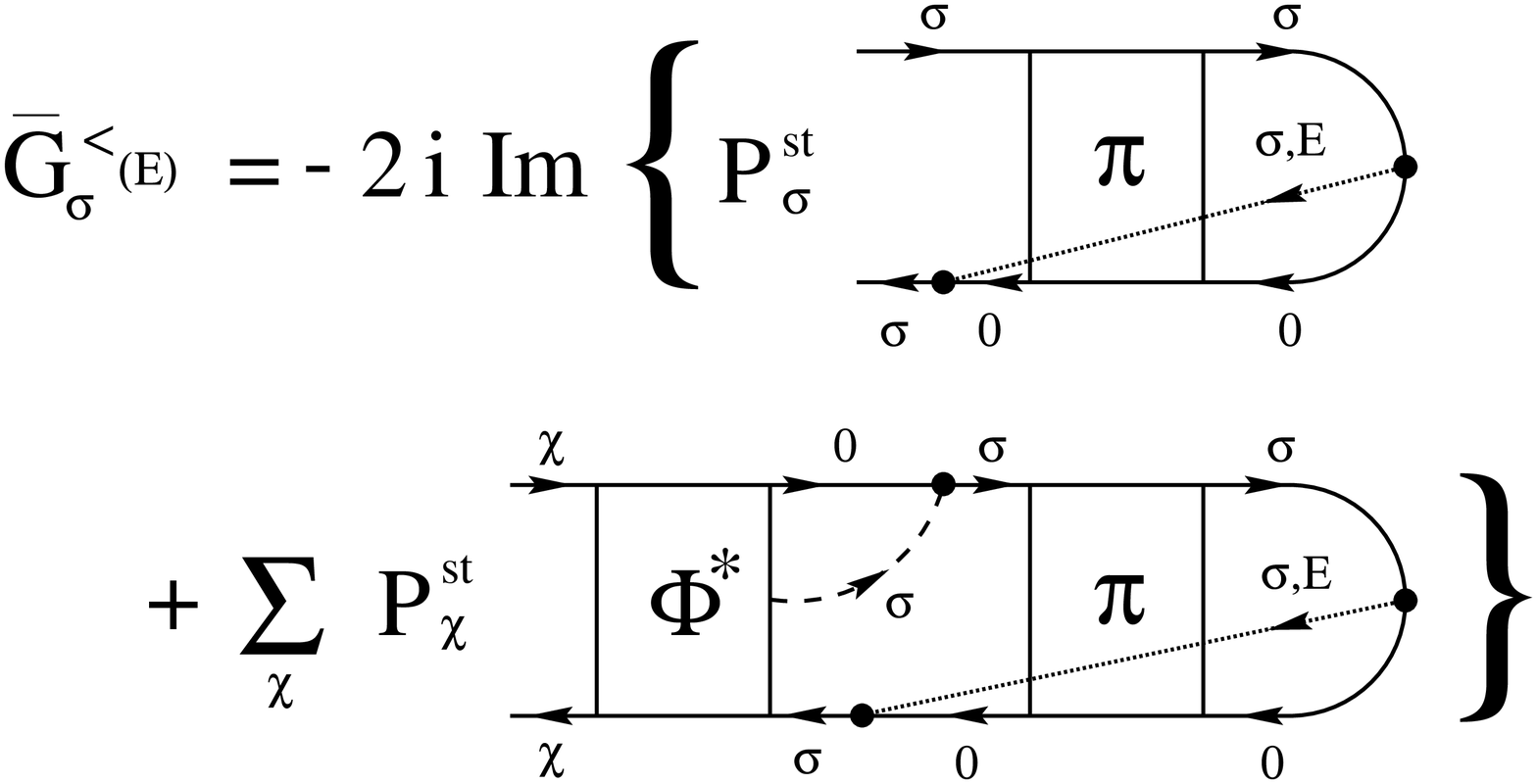,height=5cm}}
\caption{Graphical representation of $\bar{G}_\sigma^<(E)$.}
\label{Cdotkl}
\end{figure}

The correlation functions can be calculated from the diagrams shown in
Figs.~\ref{Cdotgr} and \ref{Cdotkl}.
We have to consider only the latest (i.e. rightmost) correlated part of the 
diagram.
The processes before end up with probability $P^{st}_{\chi}$ in a diagonal 
state $\chi$.
We have used the same criterion as for the calculation of the density matrix 
with one exception. 
If a vertical line lies between the external vertices we allow a cut through 
at most one tunneling line. 
Here we have used the fact that such a vertical line will in addition always 
cut the virtual line connecting the external vertices. 
The sum of all these diagrams gives (where we can combine always two diagrams 
to the imaginary part of one of them)
\begin{eqnarray}
	G_\sigma^>(E)&=& \int dE'\, \bar{G}_\sigma^>(E') P^+(E'-E)
	\\
	G_\sigma^<(E)&=& \int dE'\, \bar{G}_\sigma^<(E') P^-(E'-E)
\end{eqnarray}
with
\begin{eqnarray}\label{correlation function1}
	\bar{G}_\sigma^>(E)&=&2i\,\mbox{Im}\left\{\pi^\sigma(E) \left[
	P_0^{st} - 
	\sum_\alpha \sum_{\sigma'} \int \, dE'{P_0^{st}
	{{\phi^*}^{0,\sigma'}_{0,0}(\alpha,\sigma',E')}
	+\sum_{\sigma''}P_{\sigma''}^{st}
	{{\phi^*}^{\sigma'',\sigma'}_{\sigma'',0}(\alpha,\sigma',E')}
	\over E-E'+\epsilon_{\sigma'}-\epsilon_\sigma+i0^+} \right]\right\}
	\\
	\label{correlation function2}
	\bar{G}_\sigma^<(E)&=&-2i\,\mbox{Im}\left\{\pi^\sigma (E) \left[
	P_{\sigma}^{st} +
	\sum_\alpha \int \, dE'{P_0^{st}{\phi^*}^{0,\sigma}_{0,0}
	(\alpha,\sigma,E')
	+\sum_{\sigma'}P_{\sigma'}^{st}{\phi^*}^{\sigma',\sigma}_{\sigma',0}
	(\alpha,\sigma,E')
	\over E-E'+i0^+} \right] \right\}\, .
\end{eqnarray}

In the following, we discuss for transparency the effect of the 
coupling to bosons and the presence of a magnetic field separately.

\subsection{Boson-assisted tunneling}

For zero magnetic field, i.e. $\epsilon_{\sigma}=\epsilon$ for all $\sigma$, 
we can perform the resummation of the corresponding diagrams for the rates and
the Green's functions analytically (details are given in 
Appendix~\ref{append boson}) and find
\begin{equation}\label{current boson}
	I^{st}_\alpha = 2\pi e M \sum_{\alpha'}\int dE \,
	[\gamma_\alpha^-(E)\gamma_{\alpha'}^+(E) - \gamma_{\alpha'}^-(E)
	\gamma_\alpha^+(E)]\,|\pi (E)|^2\,
\end{equation}
with $\pi (E)=\pi^\sigma (E)$.
We can write this equation in a more intuitive way by inserting the 
definition (\ref{gamma}) for $\gamma^\pm_\alpha$ 
\begin{equation}\label{transmission formula}
	I_\alpha^{st}=\frac{e}{h}\sum_{\alpha'}\int dE \int dE'
	\left\{ T_{\alpha',\alpha}(E',E) f_{\alpha'}(E')(1-f_\alpha (E))-
	T_{\alpha,\alpha'}(E,E') f_\alpha (E) (1-f_{\alpha'}(E'))\right\}\,,
\end{equation}
where
\begin{equation}\label{transmission probability}
	T_{\alpha,\alpha'}(E,E')=M\Gamma_\alpha (E)\Gamma_{\alpha'}(E')
	\int dE_1 P^+(E_1-E)P^-(E_1-E') |\pi(E_1)|^2
\end{equation}
can be interpreted as a transmission probability for an electron to start from
reservoir $\alpha$ with energy $E$ and end in reservoir $\alpha'$ with energy 
$E'$. 
From the detailed balance condition (\ref{detailed balance}) we get
\begin{equation}\label{trans detailed balance}
	T_{\alpha',\alpha}(E',E) = e^{\beta (E'-E)}
	T_{\alpha,\alpha'}(E,E')\,.
\end{equation}
This guarantees that the current is zero if all chemical potentials of the 
reservoirs are identical.

However, the interpretation of $T_{\alpha,\alpha'}$ as a one-particle
transmission probability in analogy to a generalization of the
Landauer-B\"uttiker formula to inelastic interactions \cite{Hek-Naz-Sch} is 
not correct. 
We see that the transmission probability still depends on the Fermi 
distribution functions via the self-energy $\sigma (E)$ in the denominator of 
the propagator $\pi (E)$. 
This reflects the many-particle aspect of the electron-electron and 
electron-boson interaction in our model.

Comparing our result for $T_{\alpha,\alpha'}$ with other approaches in the 
case $M=1$ \cite{Win-Jac-Wil,Gla-She,Jon,Ima-Pon-Ave}, we see that the energy 
dependence of $\sigma(E)$ has been neglected in all previous treatments.
We find that even in the $M=1$ case, the energy dependence of $\sigma(E)$ 
cannot been neglected if the temperature $T$ and the typical frequency 
$\omega_B$ of the bosons are smaller than $\Gamma$. 

Without bosons, the current formula is exact up to order $O(\Gamma^2)$, i.e., 
sequential and electron cotunneling are fully taken into account. 
With bosons, cotunneling is not described correctly since we have treated the 
bosons only by a dressing of the tunneling lines. 
This means that our approximation is not valid in regions where the current is
very small. 
However, at resonance we believe our treatment to be correct since there we 
expect that sequential tunneling will be just modified by a renormalization 
and broadening of the local state of the dot which is described by the 
self-energy $\sigma (E)$ which is calculated in lowest order in $\Gamma$ here.
Higher orders will be small for high tunneling barriers.

Finally, we calculate the Green's functions and find
\begin{eqnarray}
	G^>(E)&=&-2\pi i \int dE'\,\gamma^-(E')P^+(E'-E)\,|\pi(E')|^2\\
	G^<(E)&=&2\pi i \int dE'\,\gamma^+(E')P^-(E'-E)\,|\pi(E')|^2\,.
\end{eqnarray}
In equilibrium, i.e. $\mu_\alpha = 0$ for all $\alpha$, we obtain the correct 
sum rule $G^>(E)=-\exp{(\beta E)}G^<(E)$. 
Furthermore, for the $M=1$ case, particle-hole symmetry is satisfied.
The spectral density has the form
\begin{equation}
	\rho(E)=\int dE' \left[ \gamma^+(E') P^-(E'-E) +
	\gamma^-(E') P^+(E'-E) \right] |\pi(E')|^2\, .
\label{spec}
\end{equation}
The effect of the resonant-tunneling processes is described by the resolvent 
$\pi(E)$ containing the self-energy $\sigma(E)$, Eq.~(\ref{sigma2}).
The real and imaginary part of the self-energy express energy renormalization 
and broadening and determine, therefore, the position and the width of the 
maxima in the spectral density.

To proceed we consider from now on a one-mode environment (Einstein model) 
with boson frequency $\omega_q=\omega_B$. 
Experimentally realizations of this model are optical phonons 
\cite{Win-Jac-Wil,Gla-She,Jon} or by fluctuations of an external $LC$-circuit 
with frequency \cite{Devoret-etal,Odi-etal,Flens-etal,Ima-Pon-Ave}
$\omega_B=(LC)^{-1/2}$.
The results for a general environment can be anticipated approximately from 
the one-mode case by a superposition.
Furthermore, we choose the special case of two reservoirs $\alpha=L/R$ and 
constant level broadening $\Gamma/2=\Gamma_L=\Gamma_R$.

Defining $g=\sum_q {g_q^2 / \omega_B^2}$ we obtain 
$P^\pm(E)=\sum_n p_n \delta(E\pm n\omega_B)$, where
$p_n=e^{-g(1+2N_0(\omega_B))}e^{n\omega_B/2T_B} I_n(2gN_0(\omega_B) 
e^{\omega_B/2T_B})$
is the probability for the emission of $n$ bosons with frequency $\omega_B$.
Here, $N_0(\omega_B)$ is the Bose function and $I_n$ the modified Bessel 
function.
Using Eq.~(\ref{sigma2}) we obtain \cite{Koe-HS-Sch2}
\begin{equation}
	\mbox{Re}\,\sigma(E)=\sum_{n,\alpha} (M p_n - p_{-n}){\Gamma_{\alpha}
	\over 2\pi}
        \Big{[}  \ln \left( {E_C\over 2\pi T} \right)
	-\mbox{Re}\,\Psi \left({1\over 2}
	+i{E+n\omega_B-\mu_\alpha \over 2\pi T} \right) \Big{]}
\label{Resigma}	
\end{equation}
\begin{equation}
	\mbox{Im}\, \sigma(E)=-\pi\sum_n p_n[M \bar{\gamma}^+(E+n\omega_B)+ 
	\bar{\gamma}^-(E-n\omega_B)] \, .
\end{equation}
Here $\Psi$ denotes the digamma function, and we have chosen in the energy 
integrals a Lorentzian cut-off at $E_C$.

The real part of $\sigma(E)$ renormalizes the level position to higher 
energies.
Furthermore, it depends logarithmically on energy, temperature, voltage and 
frequency. 
These logarithmic terms are typical for the occurrence of Kondo peaks.
Hence, we anticipate logarithmic singularities either for $M\ge 2$ or for 
$p_n\neq p_{-n}$.
This includes not only the degenerate case but also the case of a single dot 
level without spin ($M=1$) since the probabilities for absorption and 
emission of bosons are different. 
It is important to remark here, that for systems coupled to classical 
time-dependent fields\cite{Het-HS} the situation is different since then both 
probabilities are equal.
At low enough temperatures we obtain logarithmic peaks in $\sigma(E)$ at 
$E=\mu_\alpha + n\omega_B$ ($n\neq 0$ for $M=1$).
They lead to maxima of the resolvent $\pi(E)$ at $E=\mu_\alpha + n\omega_B$
($n>0$ for $M=1$, $n\geq 0$ for $M>1$) for $\epsilon<0$ and at 
$E=\mu_\alpha + n\omega_B$ ($n<0$) for $\epsilon>0$.
The spectral density (\ref{spec}) shows resonances at the same points but, 
due to the additional $P^\pm$ functions in the integrand, they are shifted by 
multiples of $\omega_B$.
This boson-assisted tunneling is completely independent from the influence of 
the bosons on the self-energy $\sigma(E)$.

\begin{figure}
\centerline{\psfig{figure=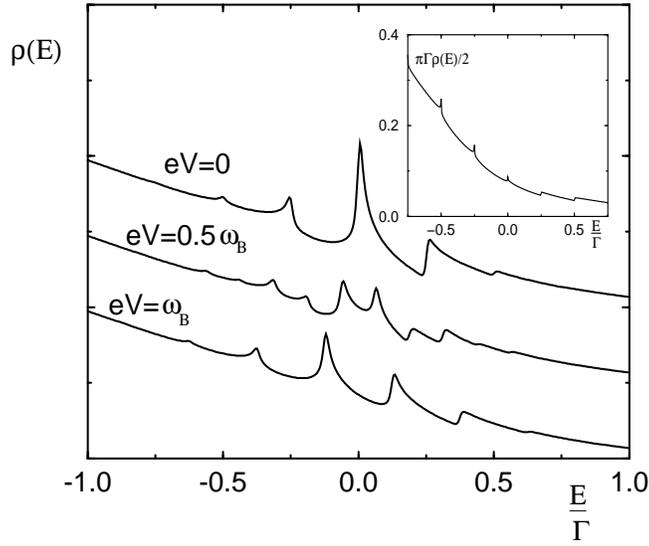,height=8cm}}
\caption{The spectral density for $M=2$, $T=T_B=0.005\Gamma$, 
	 $\epsilon=-2\Gamma$, $g=0.2$, $\omega_B=0.25\Gamma$ and 
	 $E_C=50\Gamma$ at different voltages.
         For $V=0$ there are resonances at multiples of $\omega_B$, which 
	 split for finite bias voltage.
	 Inset: spectral density for $M=1$, $T=0.00005\Gamma$, 
	 $T_B=0.5\Gamma$, $\epsilon=-\Gamma$, $V=0$, $g=0.5$, 
	 $\omega_B=0.25\Gamma$ and $E_C=50 \Gamma$.}
\label{fig2}
\end{figure}

The spectral density at different voltages for a low lying level 
$\epsilon < 0$ is depicted in Fig.~\ref{fig2}. 
Without an applied bias voltage, we obtain (for $M=2$) the usual Kondo peak 
near the Fermi level (which we choose as zero energy).
The emission of bosons leads to additional resonances at multiples of 
$\omega_B$. 
For $M=1$ and $\epsilon<0$, resonances occur for negative energies and in the 
case $\epsilon>0$ we find resonances at positive energies. 
In these cases, the effects are less pronounced and are only visible for very 
low temperatures.
At finite bias voltages all peaks split and decrease in magnitude. 

\begin{figure}
\centerline{\psfig{figure=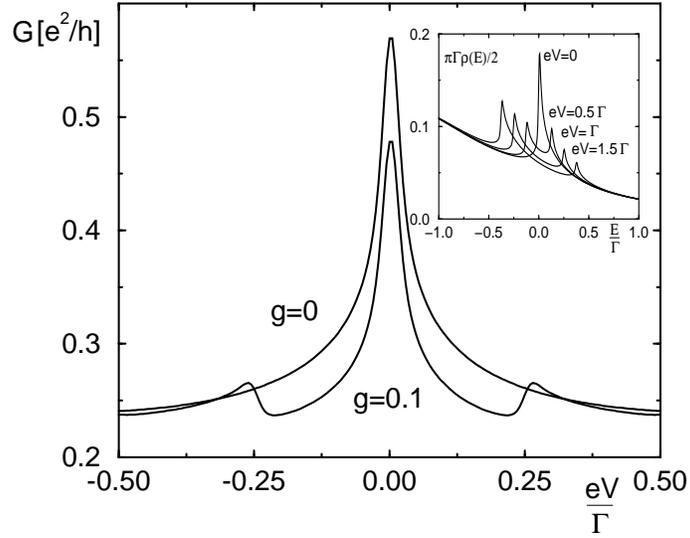,height=8cm}}
\caption{The differential conductance vs. bias voltage for 
         $T=T_B=0.005\Gamma$, $\epsilon=-2\Gamma$, $\omega_B=0.25\Gamma$ and 
         $E_C=50\Gamma$. 
	 The curves show a maximum at zero bias and satellite maxima at 
	 multiples of $\omega_B$ for a finite electron-boson coupling. 
	 Inset ($g=0$): increasing voltage leads to an overall decrease of the
 	 spectral density in the range $|E|<eV$.}
\label{fig3}
\end{figure}

The resonances in the spectral density can be probed by the nonlinear
differential conductance as function of the bias voltage $V$, as shown in 
Fig.~\ref{fig3} for the case $\epsilon < 0$. 
The splitting of the Kondo peak leads to an overall decrease of the 
spectral density in the energy range $|E|<eV$ (see inset of Fig.~\ref{fig3}). 
For this reason, the conductance shows the well-known
\cite{Mei-Win-Lee,Het-HS,Ral-Buh1} maximum at zero bias.
The emission of bosons produces a set of symmetric satellite maxima. 
They can be traced back to the fact that pairs of Kondo peaks can merge if the
bias voltage is a multiple of the boson frequency (see Fig.~\ref{fig2}). 
This gives rise to pronounced Kondo peaks at $E=\pm eV/2$ and thus to an 
increase of the spectral density with bias voltage near these points.

\begin{figure}
\centerline{\psfig{figure=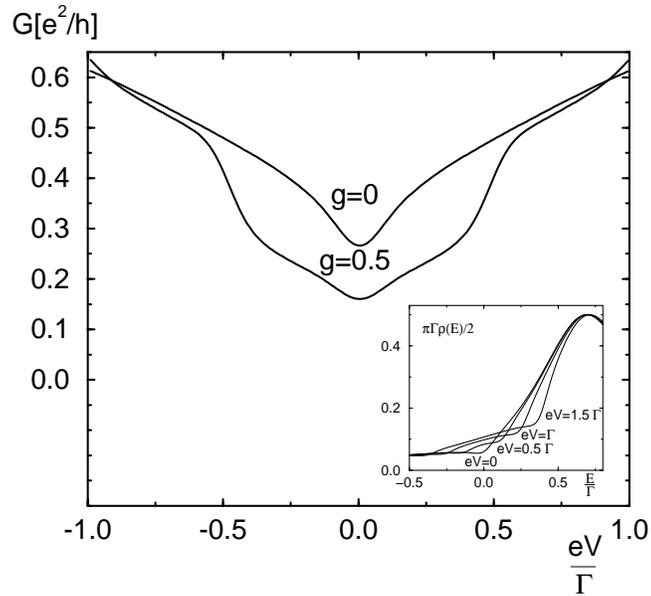,height=8cm}}
\caption{The differential conductance vs. bias voltage for 
	 $T=T_B=0.025\Gamma$, $\epsilon=0$, $\omega_B=0.5\Gamma$ and 
	 $E_C=50\Gamma$. 
	 The curves show a minimum at zero bias and steps at multiples of 
	 $\omega_B$ for a finite electron-boson coupling. 
	 Inset ($g=0$): increasing voltage leads to an overall increase of
 	 the spectral density in the range $|E|<eV$.}
\label{fig4}
\end{figure}

Fig.~\ref{fig4} shows the differential conductance for $\epsilon \ge 0$ with 
and without bosons. 
A striking result is that the whole structure is inverted compared to the case
$\epsilon<0$, and we find a zero-bias anomaly although the Kondo peak at zero 
energy is absent.
The coupling to bosons yields satellite steps at $|eV|=n\omega_B$. 
The contributions of sequential and cotunneling lead, compared to resonant 
tunneling, only to a weak bias voltage dependence of the differential 
conductance.
This shows clearly that the influence of the logarithmic terms in $\sigma(E)$ 
are still important.
The logarithmic peaks in $\mbox{Re}\, \sigma(E)$ decrease with increasing bias
voltage and approach the value of $E-\epsilon$ if $\epsilon$ is large enough. 
Thus the value of $E-\epsilon-\mbox{Re}\, \sigma(E)$ decreases which in turn 
leads to an overall increase of the resolvent $\pi(E)$ and the spectral 
density $\rho(E)$ near zero energy (see the inset of Fig.~\ref{fig4}). 
 
Zero-bias minima are known from Kondo scattering from magnetic impurities.
\cite{Jan-etal}
They have been observed in recent experiments \cite{Ral-Buh2} and have been 
interpreted as 2-channel Kondo scattering from atomic tunneling systems
\cite{Ral-Lud-Del-Buh,Het-Kro-Her} or by tunneling into a disordered metal. 
\cite{Win-Alt-Mei}
Here we have shown that zero-bias minima can also arise due to resonant 
tunneling via local impurities if the level position is high enough such that 
we are in the mixed valence regime. 

\begin{figure}
\centerline{\psfig{figure=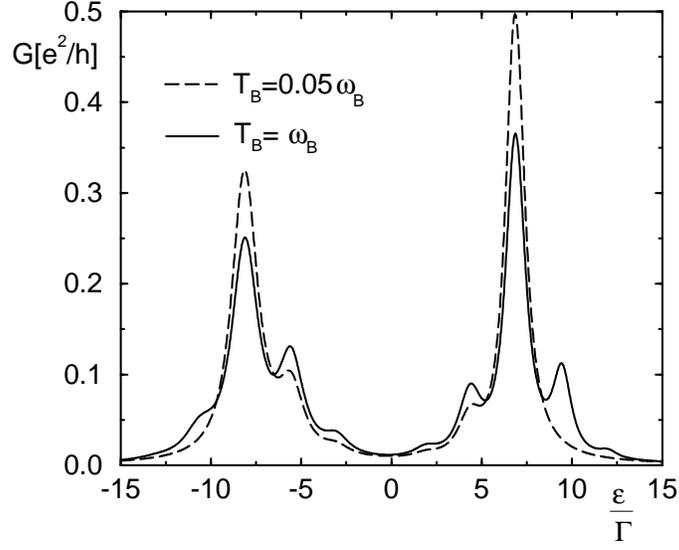,height=8cm}}
\caption{The differential conductance as a function of $\epsilon$ for
         $T=0.125\Gamma$, $eV=15\Gamma$, $g=0.3$, $\omega_B=2.5\Gamma$ and 
         $E_C=250\Gamma$.}
\label{fig5}
\end{figure}

Finally, we have investigated the differential conductance at fixed bias 
voltage as function of the position of the dot level, which experimentally
can be varied by a gate voltage coupled capacitively to the dot 
(see Fig.~\ref{fig5}).
The result shows a (classical) pair of peaks at $|\epsilon|=eV/2$ together 
with satellites (due to emission and absorption of bosons) and peaks for 
$|\epsilon|>eV/2$ (only due to absorption). 
The energy dependence of $\mbox{Im}\, \sigma(E)$ gives rise to an asymmetry of
the peak heights. 
The peak at $\epsilon=eV/2$ is higher than the one at $\epsilon=-eV/2$ since
$|\mbox{Im}\, \sigma(E)|=\pi |M\gamma^+(E)+\gamma^-(E)|$ is smaller for higher
energies (except for $M=1$ when particle-hole symmetry holds). 
This significant effect is due to the broadening of the spectral density by 
quantum fluctuations.

\subsection{Magnetic field dependence}

In this section we discuss the effect of an applied magnetic field and do not 
take into account the coupling to bosons.
Again we consider the case of two reservoirs and constant level broadenings.
Since the energy levels $\epsilon_{\sigma}$ are now spin dependent, we can no 
longer solve the self-consistent equations analytically but have to solve them
numerically.

We find Kondo resonances in the spectral density $\rho_{\sigma}(E)$ at 
energies $E=\mu_{\alpha}+\epsilon_{\sigma'}-\epsilon_{\sigma}$ with 
$\sigma' \neq \sigma$.
This is due to the fact, that the correlation functions $G^<_\sigma(E)$ and
$G^>_\sigma(E)$ are mainly determined by the resolvent $\pi^\sigma(E)$ 
(see Eqs.~(\ref{correlation function1}) and (\ref{correlation function2})) 
which contains via the self-energy $\sigma^\sigma(E)$ logarithmic 
singularities at the corresponding energies, 
\begin{eqnarray}
	\mbox{Re}\,\sigma^\sigma(E)&=&\sum_{\alpha}{\Gamma_{\alpha}\over 2\pi}
        \sum_{\sigma' \neq \sigma} \Big{[}  \ln \left( {E_C\over 2\pi T} 
	\right) -\mbox{Re}\,\Psi \left({1\over 2}
	+i{E+\epsilon_{\sigma'}-\epsilon_\sigma-\mu_\alpha \over 2\pi T} 
	\right) \Big{]}
	\\
	\mbox{Im}\, \sigma^\sigma(E)&=&-\pi \left[ \bar{\gamma}^-(E)
	+ \sum_{\sigma'} \bar{\gamma}^+(E+\epsilon_{\sigma'}-\epsilon_\sigma)
	\right] \, .
\end{eqnarray}

\begin{figure}
\centerline{\psfig{figure=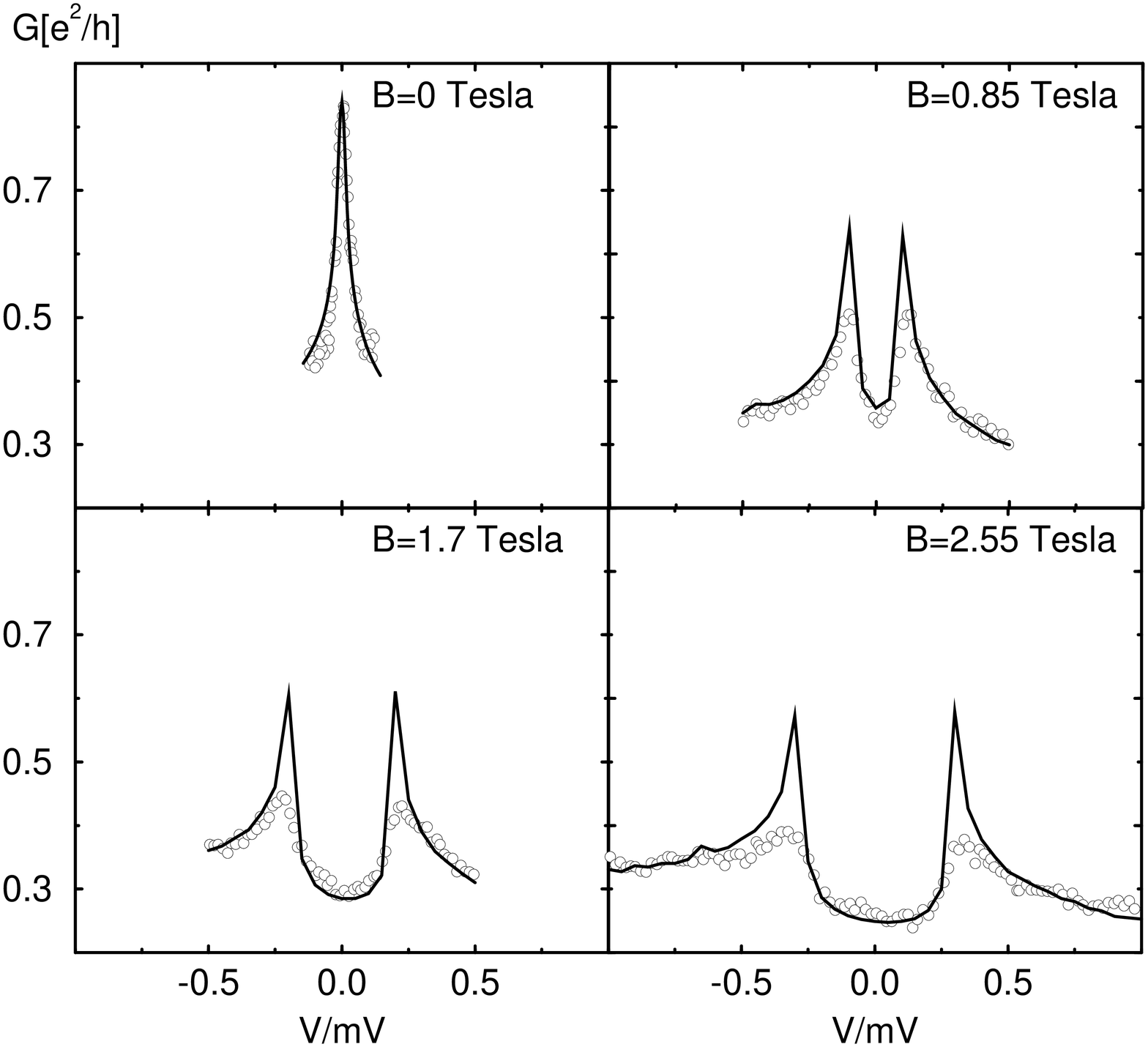,height=8cm}}
\caption[A]{The differential conductance vs. bias voltage for $T=4.3\mu eV$,
	$\epsilon_\sigma(B=0)=-5.2meV$, $\Gamma=3.4meV$,
	$a_c=0.33$, and $E_C=30meV$. 
	The circles are experimental data from Ref.~\onlinecite{Ral-Buh1}.}
\label{Fig.experiment}
\end{figure}

From Eq.~(\ref{strom}) we see that only energies within the window defined by 
the difference of the Fermi functions contribute to the current.
For this reason, there is no Kondo-assisted tunneling at low transport voltage
but sets on if transport voltage and level splitting are equal. 
Therefore, for low lying levels the conductance peak at zero bias found in the
previous section now splits up into two peaks separated by the twice the level
splitting \cite{Mei-Win-Lee} (see Fig.~\ref{Fig.experiment}).

Ralph and Buhrman recently measured Kondo-assisted tunneling via a single 
charge trap of a point contact tunnel barrier.\cite{Ral-Buh1}
We follow the model proposed by the authors interpretating the experiment as
a realization of the Anderson model with strong Coulomb repulsion such that
double occupancy does not occur.
However we think that the interaction energy $U$ and not the conduction band 
width is the relevant cut-off in this situation.

\begin{figure}
\centerline{\psfig{figure=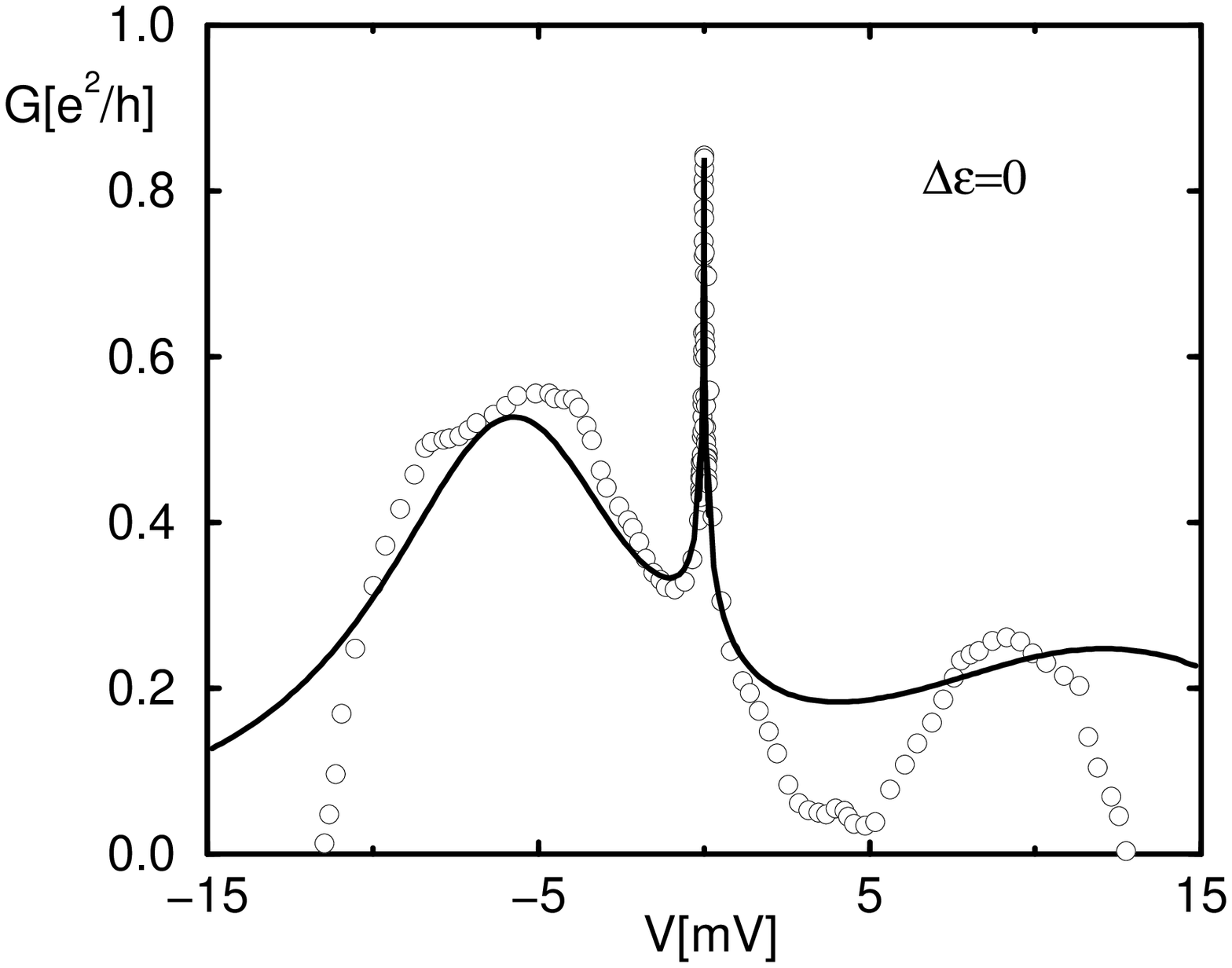,height=8cm}}
\caption[A]{The differential conductance vs. bias voltage for $T=4.3\mu eV$,
	$\epsilon_\sigma=-5.2meV$, $B=0$, $\Gamma=3.4meV$,
	$a_c=0.33$, and $E_C=30meV$. 
	The circles are experimental data from Ref.~\onlinecite{Ral-Buh1}.}
\label{Fig.expgross}
\end{figure}

\begin{figure}
\centerline{\psfig{figure=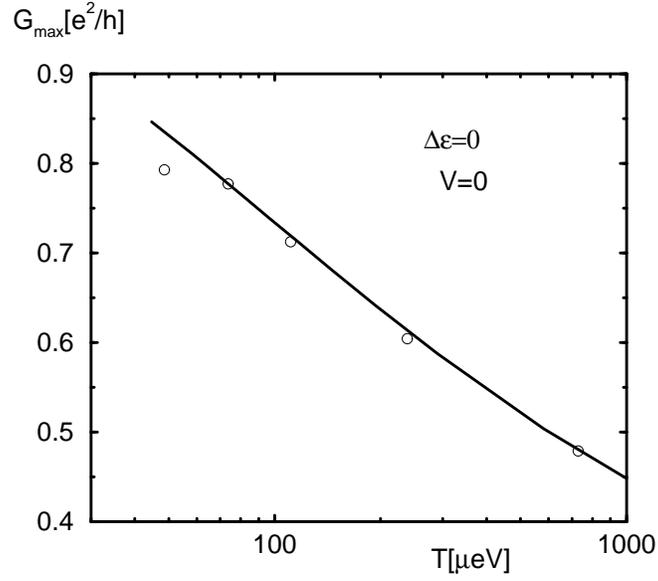,height=8cm}}
\caption[A]{The maximal linear conductance vs. temperature for
	$\epsilon_\sigma=-5.2meV$, $B=0$, $\Gamma=3.4meV$,
	$a_c=0.33$, and $E_C=30meV$. 
	The circles are experimental data from Ref.~\onlinecite{Ral-Buh1}.}
\label{Fig.exptemp}
\end{figure}

A comparison of the experiment and our theory is given in 
Fig.~\ref{Fig.experiment}, Fig.~\ref{Fig.expgross} and Fig.~\ref{Fig.exptemp}.
We find a good agreement for the peaks induced by Kondo-assisted tunneling 
processes if we set the cut-off $U=30meV$.
The authors suspect the single-charge trap to be a dangling bond, for which 
they expect $U=100meV$. 
Our result agrees in the order of magnitude, it gives a hint however, that the
state may have a larger extension than an ordinary dangling bond or that there
is screening due to the copper electrodes or both. 
The peaks for larger magnetic fields show, however, a stronger broadening than
predicted from our calculation. 
Nevertheless, our theory reproduces the experimental curves much better than
the fits given in Ref.~\onlinecite{Ral-Buh1} using perturbation theory since 
we have taken into account nonperturbative effects which are obviously 
important here.

The model proposed by the authors of Ref.~\onlinecite{Ral-Buh1} explains the 
broad peaks at large voltages by the matching of the energies of the empty and
the singly occupied dot.
Our calculations for this case, however, lead to a broader and lower peak
for positive voltages in comparison with experiment 
(see Fig.~\ref{Fig.expgross}).
We think, therefore, that due to the capacitance asymmetry the system becomes 
doubly occupied before the empty state is energetically favorable.
The capacitance asymmetry $a_c$ makes then the corresponding resonance peak 
sharper.
An energy dependent transparency of the barriers could then explain the
different heights.
A generalization of our theory to situations, where multiple occupancy of the
dot is important, is currently under way and will be presented elsewhere.

\begin{figure}
\centerline{\psfig{figure=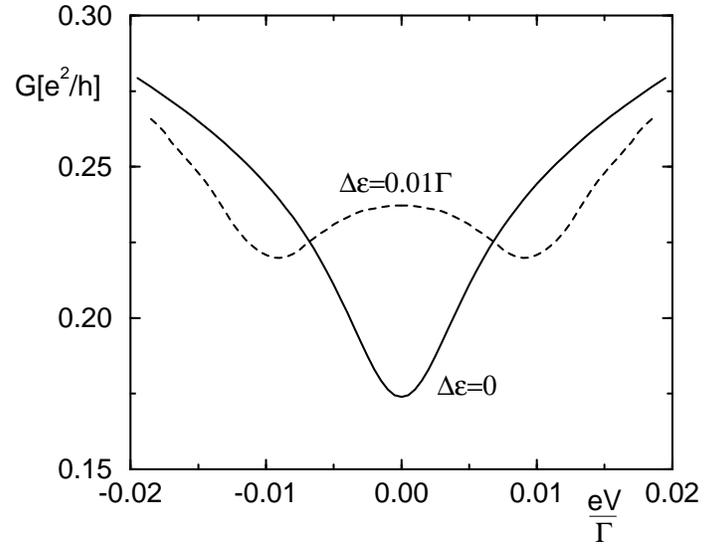,height=8cm}}
\caption{The differential conductance vs. bias voltage for $T=0.001\Gamma$,
	$\epsilon_\sigma=0.1\Gamma \pm \Delta\epsilon /2$, and 
	$E_C=10\Gamma$.}
\label{Fig.minimum}
\end{figure}

Finally, we consider the case when the energy level is above the Fermi 
energies of the leads. 
The zero-bias minimum found in the previous section splits for finite magnetic
field into two minima separated by twice the level splitting 
(see Fig.~\ref{Fig.minimum}).

\section{Conclusions}

In conclusion, we have studied low-temperature transport in the nonequilibrium
Anderson model with bosonic interactions. 
The latter yield new Kondo resonances in the spectral density which can be 
probed by the measurement of the nonlinear differential conductance. 
Both the gate and bias voltage dependence are important. 
Quantum fluctuations due to resonant tunneling yield zero-bias anomalies as 
function of the bias voltage, {\it which can be changed from maxima to minima 
by varying the gate voltage}. 
We, furthermore, discussed the splitting of the zero-bias anomaly by an 
external magnetic field and found good agreement with recent experiments.

We have presented a real-time approach which is based on a nonperturbative 
calculation of transition rates between different states of a local strongly 
correlated system coupled to fermionic or bosonic baths. 
We present systematic rules of how to set up well defined perturbation 
expansions for the rates in terms of the tunneling matrix elements between dot
and leads.
The formally exact rates are used to calculate occupation probabilities and 
the current from Master equations and current formulas which are intuitively 
obvious.
The method has a wide applicability, ranging from the study of arbitrary dot 
level structures up to the investigation of macroscopic quantum coherence 
phenomena. 
The latter can arise from the time evolution of non-stationary initial states 
or by the application of explicitly time-dependent fields. 

The usage of real-time methods to understand low-temperature behavior of
strongly correlated fermions either in equilibrium or non-equilibrium 
situations is a rather new field and has not yet been applied extensively. 
Compared with the conventional methods in imaginary time \cite{Kei-Mor} they 
offer the possibility to set up new approximation schemes.
In this paper we have performed a nonperturbative resummation of higher order 
coherent tunneling processes to calculate transition and current rates 
analytically for temperatures smaller than the intrinsic broadening $\Gamma$. 
Although the criterium for considering certain diagrams is yet not motivated
by the usage of a ``small'' parameter, the diagrams are selected in a 
systematic way.
We have chosen all diagrams which keep the total density matrix as close as
possible to the diagonal state up to one electron-hole pair excitation in the 
reservoirs. 
This reminds of techniques applied within variational wave function ansatzes
\cite{Sch-etal} but here formulated on the basis of density matrices for 
nonequilibrium systems at finite temperatures.
Furthermore, there are many possibilities to improve our approximation by 
considering more diagrams by analytical or numerical methods. 
Simple limiting cases as e.g. the noninteracting case are already exactly 
incorporated within our approximation. 
Since the strongly interacting case gives also at least qualitatively good 
results, our method may be a good candidate to cover the whole range from weak
to strong interaction within the same approximation scheme. 

We like to thank D. Averin, J. von Delft and M. Hettler for useful 
discussions. 
The support by the Deutsche Forschungsgemeinschaft, through SFB 195, by the 
Swiss National Science Foundation (H.S.), and by the A.v.Humboldt award of the
Academy of Finland (G.S.) is gratefully acknowledged.

\begin{appendix}

\section{The rules in time space}
\label{append rules in time}
Each term of the expansion Eq.~(\ref{expansion}) with operators $A_i$ of the
form Eq.~(\ref{special Ai}) can be calculated according to the following 
rules:
\begin{description}
\item[  1.  ]
Draw all topological different diagrams with directed tunneling lines 
connecting pairs of internal or external vertices containing lead electron
operators. 
Assign a reservoir index $\alpha$ and a spin index $\sigma$ to each of these 
lines.
Connect all vertices containing boson operators in all possible ways by boson 
lines.
Assign states $\chi$ and the corresponding energy $\epsilon_{\chi}$ to each 
element of the Keldysh contour connecting two vertices.
\item[  2.  ]
The propagation from $t'$ to $t$ with $t'<t$ on the Keldysh contour implies a 
factor $\exp[- i \epsilon_{\chi}(t-t')]$. 
\item[  3.  ]
The state $\chi$ which is assigned to the left most part of the diagram 
implies a factor $P_\chi^0$ from the initial density matrix. 
Each vertex containing a dot operator $B$ gives rise to a matrix element 
$\langle \chi '| B |\chi\rangle$ where $\chi$ ($\chi '$) is the dot state 
entering (leaving) the vertex with respect to the Keldysh contour.
\item[  4.  ]
Each directed tunneling line with index $\alpha$ running from $t'$ to $t$ 
implies $(-1)^v\bar{\gamma}^K_{\alpha}(t,t')$ with $v$ being the number
of electron operators (due to external vertices) on the part of the Keldysh
contour from $t'$ to $t$.
The line corresponds to a tunneling process in reservoir $\alpha$. 
Each boson line connecting vertices at times $t$ and $t'$ implies $P^K(t,t')$
if the phase factors at these vertices have different sign.
Otherwise, the boson line has the value $P^K (t,t')^{-1}$.
\item[  5.  ]
Each diagram carries a prefactor $(-i)^m(-1)^c$, where $m$ is the total number
of internal vertices and $c$ the number of crossings of tunneling lines.
There may be a further minus sign due to the order of dot electron operators
which emerges from the matrix elements $\langle \chi '| B | \chi\rangle$
discussed in rule 3.
\item[  6.  ]
Integrate over the internal times along the Keldysh contour without changing 
their ordering and sum over the reservoir and spin indices.
\end{description}

We emphasize that these diagrammatic rules hold for arbitrary dot Hamiltonians
$\bar{H}_D = \sum_\chi \epsilon_\chi |\chi\rangle\langle \chi|$, i.e.,
the states $\chi$ can be many-body eigenfunctions of $\bar{H}_D$ containing 
complicated correlations due to Coulomb interaction, magnetic fields, 
geometric setups, etc. .
Such eigenfunctions have been calculated for special situations 
\cite{Pfa-etal,Hae-etal} and can be used as an input for our diagrammatic 
language. 
In this paper, however, we will concentrate ourselves on the dot Hamiltonian 
(\ref{dot}) where the states $\chi$ are trivially known. 
For this special case, the matrix elements $\langle \chi '| B |\chi\rangle$
from rule 3 can only give rise to minus signs whereas they can have a more 
pronounced influence in more general situations \cite{Pfa-etal,Hae-etal}.

Furthermore, we note that the same diagrammatic rules even hold for arbitrary 
time-dependent dot Hamiltonians $\bar{H}_D(t)$ which are not diagonal in the
states $\chi$. 
In this case one has to assign two states $\chi'$ and $\chi$ to the beginning 
and the end of each element of the Keldysh contour, respectively. 
The factor $\exp[- i \epsilon_{\chi}(t-t')]$ from rule 2 is then replaced by 
the matrix element $\langle \chi | U_D(t,t')|\chi '\rangle$ where $U_D$ 
denotes the time evolution operator of $\bar{H}_D$ and $t$ ($t'$) are the 
times at the end (beginning) of the element of the Keldysh contour.

\section{The rules in energy space}
\label{append rules in energy}

We obtain the diagrammatic rules in energy space by expanding the expectation
value in Eq.~(\ref{correlation}) and then performing the time integrals.
We order the times of all internal ($m$) and external vertices ($n$) from 
left to right and label them by $\tau_j$ with $j=1,2,\ldots,m+n$ 
(with $\tau_{m+n}=0$), irrespective on which branch they are. 
The Keldysh contour integrals are now written as ordinary integrals.
This includes a minus sign for each internal vertex on the backward 
propagator.
If the initial density matrix is diagonal we then encounter expressions of the
type
\begin{eqnarray}\label{integral}
	\int^0_{-\infty} d\tau_1 \int^0_{\tau_1} d\tau_2
	\ldots \int^0_{\tau_{m+n-2}} d\tau_{m+n-1} \, e^{0^+ \tau_1}
	e^{-i \Delta E_1(\tau_1-\tau_2)} 
	e^{-i \Delta E_2(\tau_2-\tau_3)} \cdots
	e^{-i \Delta E_{m+n-1}\tau_{m+n-1}}
	\nonumber \\
	=\, i^{m+n-1} \frac{1}{\Delta E_1+i0^+}\, \cdot \, 
	\frac{1}{\Delta E_2+i0^+}
	\cdots \frac{1}{\Delta E_{m+n-1}+i0^+} \; .
\end{eqnarray}
Here $\Delta E_j$ is the difference of all energies going to the left minus 
all energies going to the right in each segment limited by $\tau_j$ and 
$\tau_{j+1}$.
This includes the energies of the propagators and -- if present -- the 
energies of the tunneling, boson and virtual lines.
The convergence factor $e^{0^+ \tau_1}$ is related to an adiabatic switching 
on of the tunneling term $\bar{H}_T$. 
The factor $i^{m+n-1}$ cancels with the factor $(-i)^m$ from rule 4 above 
together with the prefactor $(-i)^{n-1}$ from the definition 
Eq.~(\ref{integral}).
Therefore, the corresponding rules in energy representation read:
\begin{description}
\item[  1'.  ]
Draw all topologically different diagrams with fixed ordering of the vertices
along the real axis, i.e., irrespective on which branch they are. 
The vertices are connected by tunneling and boson lines as in time space. 
In addition to the energy $\epsilon_{\chi}$ assigned to the propagators we 
assign an energy $E$ to each tunneling line.
For each boson line choose a direction (arbitrarily) and assign also an energy
$E$. 
The external vertices are connected by virtual lines with energies $E_i$ 
($i=2,\ldots,n$) as described above.
\item[  2'.  ]
For each segment derived from $\tau_j \le \tau \le \tau_{j+1}$ with 
$j=1,2,\ldots,m+n-1$ assign a resolvent ${1\over \Delta E_j +i0^+}$ where 
$\Delta E_j$ is the difference of the leftgoing minus the rightgoing energies 
(including the energies of the tunneling, boson and virtual lines).
\item[  3'.  ]
See rule 3 in time space.
\item[  4.'  ]
For each coupling of vertices write $(-1)^v \bar{\gamma}^+_{\alpha}(E)$, if 
the tunneling line of reservoir $\alpha$ is going backward and 
$(-1)^v \bar{\gamma}^-_{\alpha}(E)$, if it is going forward with respect to 
the closed time path (definition of $v$ see rule 4 in time space).
For each boson line write $P^+(E)$, if it is going backward and $P^-(E)$, if 
it is going forward with respect to the closed time path.
\item[  5'.  ]
The prefactor is given by $(-1)^b(-1)^c$, where $b$ is the total 
number of internal vertices on the backward propagator and $c$ the number of 
crossings of tunneling lines.
There may be a further minus sign due to the order of dot electron operators
which emerges from the matrix elements $\langle \chi'| B | \chi\rangle$
discussed in rule 3.
\item[  6'.  ]
Integrate over the energies of tunneling and boson lines and sum over the 
reservoir and spin indices.
\end{description}

\section{Current conservation}
\label{append proof}

In this appendix we prove Eq.~(\ref{rate relation}) and current conservation.
Let us consider any diagram $\Sigma^{\alpha p}_{\chi',\chi}(t',t)$ in the 
expression 
\begin{equation}\label{append rate}
	\sum_{\chi}\sum_p p\,\Sigma^{\alpha p}_{\chi',\chi}(t',t) \, .
\end{equation}
By changing the vertical position of the rightmost vertex we get a new diagram
which has up to a minus sign the same value as the old diagram from which the 
new one was constructed.
If the rightmost tunneling of the old diagram line has a reservoir index 
different from $\alpha$, then the new diagram is of the form 
$\Sigma^{\alpha p}_{\chi',\chi''}$, so that the sum of all these contributions
in Eq.~(\ref{append rate}) is zero.
The other diagrams are divided into two classes: in the one (the other) class,
the rightmost tunneling line of each diagram enters (leaves) the forward 
propagator or leaves (enters) the backward propagator.
The change of the vertical position of the rightmost vertex, then, increases 
(decreases) the value of $p$ by one, so that the new diagram is of the form
$\Sigma^{\alpha p\pm 1}_{\chi',\chi''}$. 
Furthermore, the old and the new diagram belong to different classes.
After changing the position of the rightmost vertex of only one class and then
shifting the summation index $p$ in Eq.~(\ref{append rate}), we obtain exactly
all diagrams of $\sum_{\chi}\Sigma^{\alpha +}_{\chi',\chi}$ which proves 
Eq.~(\ref{rate relation}).

The conservation of probability follows directly from the master equation 
(\ref{master}). 
Summation over $\chi$ together with $\sum_{\chi}\Sigma_{\chi',\chi}(t',t)=0$
yields
\begin{equation}\label{probability conservation}
	\sum_{\chi} \frac{d}{dt}  P_\chi (t) = 0\,.
\end{equation}

To prove current or charge conservation we first recognize that
\begin{equation}\label{conserving condition}
	\sum_{{\chi\atop N(\chi)=p}}\Sigma^{\alpha+}_{\chi',\chi}=
	-\sum_{{\chi\atop N(\chi)=p+1}}\Sigma^{\alpha-}_{\chi',\chi}
\end{equation}
where $N(\chi)$ is the particle number on the dot for state $\chi$.
This relation follows directly by changing the vertical position of the 
rightmost vertex.

After multiplication of the master equation (\ref{master}) with $-e$ and 
$N(\chi)$ and summation over $\chi$, we use Eqs.~(\ref{rate-partial rate}) and
(\ref{conserving condition}), insert the current formula (\ref{current-sigma})
and find the conservation law for the total charge flowing into the dot
\begin{equation}\label{current conservation}
	\sum_\alpha I_\alpha (t) = \frac{d}{dt} Q(t)
\end{equation}
where $Q=-eN=-e \sum_{\chi} N(\chi) P_\chi$ is the charge on the dot. 

In the stationary and time-independent case Eq.~(\ref{current conservation}) 
reduces to the conservation of the tunneling current
\begin{equation}
	\sum_\alpha I^{st}_\alpha = 0
\end{equation}
whereas for the general case the r.h.s. of Eq.~(\ref{current conservation})
is minus the sum over all displacement currents flowing in the reservoirs.

An important result of this appendix is that any approximation for the
rates is current conserving provided that the condition 
Eq.~(\ref{conserving condition}) is satisfied.
This means that we always have to consider both vertical positions of the 
rightmost vertex.

\section{Analytic solution for zero magnetic field}
\label{append boson}

For zero magnetic field, i.e. $\epsilon_{\sigma}=\epsilon$ for all $\sigma$, 
we define the quantities $\pi(E)\equiv\pi^\sigma(E)$, $\sigma(E)\equiv
\sigma^\sigma(E)$, and
\begin{equation}
	\phi^+_\alpha(E)=\phi_{0,0}^{0,\sigma}(\alpha,\sigma,E)
	\qquad \mbox{and} \qquad
	\phi^-_\alpha(E)= \sum_{\sigma'} 
	\phi_{\sigma,0}^{\sigma,\sigma'}(\alpha,\sigma',E)
\end{equation}
which are independent of $\sigma$.
We get the integral equations 
\begin{equation}\label{integral equation}
	[E-\epsilon -\sigma(E)]\phi^{\pm}_\alpha(E)=
	\pm\gamma^{\pm}_\alpha(E)-\gamma_\alpha(E)\int dE'
	{1\over E-E'+i0^+}\,{\phi^*}^{\pm}(E')
\end{equation}
where $\gamma_\alpha (E)=\gamma^-_\alpha (E) + M\gamma^+_\alpha (E)$ and 
$\phi^\pm (E)=\sum_\alpha\phi^\pm_\alpha (E)$. 
Summing over $\alpha$ and taking the imaginary part we obtain the solution
\begin{equation}\label{ImPhi}
	\mbox{Im}\,\phi^{\pm}(E)=\mp\pi{\lambda^{\pm}\over\lambda}
	\gamma(E)|\pi(E)|^2
\end{equation}
where we used the definitions 
$\gamma^\pm (E) = \sum_\alpha \gamma^\pm_\alpha (E)$,
$\gamma (E) = \gamma^- (E) + M\gamma^+ (E)$,
\begin{equation}
	\lambda^{\pm}=\int dE \, \gamma^{\pm}(E)|\pi(E)|^2
	\qquad \mbox{and} \qquad
	\lambda=\int dE \, |\pi(E)|^2 \, .
\end{equation}
Furthermore, we obtain directly from (\ref{integral equation}) a relation 
between $\phi_\alpha$ and $\phi$
\begin{equation}\label{phi-alpha}
	\gamma (E)\phi^{\pm}_\alpha (E)=\gamma_\alpha (E)
	\phi^{\pm}(E)\pm \pi (E)[\gamma (E) \gamma^\pm_\alpha (E) -
	\gamma^\pm (E) \gamma_\alpha (E)]
\end{equation}
Using (\ref{rates-special}), the current rates follow from 
$\Sigma^{\alpha +}_{0,0}=2iM\int dE \, \mbox{Im}\, \phi^+_\alpha (E)$ and
$\Sigma^{\alpha +}_{\sigma,0}=2i\int dE \, \mbox{Im}\, \phi^-_\alpha (E)\,$.
With Eqs.~(\ref{ImPhi}) and (\ref{phi-alpha}), the result is
\begin{eqnarray}\label{current rate solution}
	\Sigma^{\alpha+}_{0,0}&=&-2\pi i M \left [\frac{\lambda^+}{\lambda}
	\lambda_\alpha + \int dE \, |\pi(E)|^2 [\gamma^-(E)\gamma^+_\alpha (E)
	-\gamma^+(E)\gamma_\alpha^-(E)]\right ]\\
	\Sigma^{\alpha+}_{\sigma,0}&=&2\pi i  \left [\frac{\lambda^-}{\lambda}
	\lambda_\alpha - M \int dE\, |\pi(E)|^2[\gamma^-(E)\gamma^+_\alpha (E)
	-\gamma^+(E) \gamma_\alpha^-(E)]\right ]\,,
\end{eqnarray}
where $\lambda_\alpha = \int dE \gamma_\alpha (E) |\pi (E)|^2$.

Summing the current rates over $\alpha$ and using $\sum_\alpha \lambda_\alpha 
= \lambda^- + M\lambda^+ =1$, we get the total transition rates (note that 
$\Sigma^{\alpha-}_{\chi',0}=0$)
\begin{equation}\label{transition rate solution}
	\Sigma_{0,0}=-2\pi iM{\lambda^+\over\lambda}\qquad\mbox{and}\qquad
	\Sigma_{\sigma,0}=2\pi i{\lambda^-\over\lambda}
\end{equation}
and the solution of the stationary Master equation 
(\ref{stationary master}) reads
\begin{equation}\label{P}
	P^{st}_0=\lambda^- \qquad \mbox{and} \qquad P^{st}_{\sigma}=\lambda^+ 
	\qquad \mbox{with} \qquad \lambda^-+M\lambda^+=1 \, .
\end{equation}
The stationary current follows from (\ref{stationary current})
$I^{st}_\alpha = -ie [P_0^{st}\Sigma^{\alpha+}_{0,0}+M P^{st}_\sigma
\Sigma^{\alpha +}_{\sigma,0}]\,$
(note that $\Sigma^{\alpha+}_{\chi',\sigma}=0$), which gives as the final 
result Eq.(\ref{current boson}).

\end{appendix}

\newpage

\end{document}